\documentclass[twocolumn,trackchanges,tighten,twocolappendix]{aastex701}

\usepackage{amsmath}
\usepackage[T1]{fontenc}
\usepackage{graphicx}
\usepackage{float}

\begin{document}

\title{A pseudo-Newtonian stationary circumbinary slim disk model}

\correspondingauthor{Sixiang Wen}
\author[0000-0002-0934-2686]{Sixiang Wen}
\email{wensx@bao.ac.cn}
\affiliation{National Astronomical Observatories, Chinese Academy of Sciences, 20A Datun Road, Beijing 100101, China}
\affiliation{Department of Astrophysics/IMAPP, Radboud University, P.O.~Box 9010, 6500 GL, Nijmegen, The Netherlands}
\affiliation{Department of Astronomy, University of Arizona, Tucson, AZ 85726, USA}
\correspondingauthor{Vasileios Paschalidis}
\author[0000-0002-8099-9023]{Vasileios Paschalidis}
\email{vpaschal@email.arizona.edu}
\affiliation{Department of Astronomy, University of Arizona, Tucson, AZ 85726, USA}
\affiliation{Department of Physics, University of Arizona, Tucson, AZ 85726, USA}

\begin{abstract}

We present a pseudo-Newtonian stationary circumbinary slim disk model. We extend the slim disk formalism by including the binary tidal torque and solve the resulting steady-state equations to determine the circumbinary disk structure. We compare the binary slim disk solutions with corresponding binary thin disk solutions, calculate the disk spectrum, explore the impact of different parameters on the system, and estimate the binary shrinkage timescale. We find that; (1) due to the different disk density profiles, the integrated tidal torque exerted on the disk is significantly smaller for the slim disk than for the thin disk; as a result thin disks onto binary black holes can be radiatively significantly more efficient than slim disks; (2) The presence of the secondary alters the emission of the circumbinary disk, making it different from the spectrum of a single black hole Active Galactic Nuclei (AGN); (3) The tidal torque boosts the viscous torque in the outer part of the disk (radii greater than the binary separation), which is strongly dependent on the disk parameters, including the binary mass ratio $q$, the orbital separation $a$, the viscous parameter $\alpha$ and the accretion rate $\dot M$;  (4) The vertical component of the potential of the secondary slightly decreases the integrated tidal torque. However, both the vertical and radial components of the potential of the secondary have small impact on the disk radiative flux; (5) Using the integrated disk tidal torque backreacting on the secondary at different orbital separations, we find that the disk provides an efficient way to shrink the binary orbital separation.  

\end{abstract}


\section{Introduction} \label{sec:intro}

Accretion onto supermassive binary black holes (SMBBH) has been the topic of intense theoretical research, see, e.g., \citep{bogdanovic_bhb_review,Lai2023,Valli2024,Gutierrez2024,Cattorini2024} for reviews. The detection of gravitational wave (GW) signals by the LIGO, Virgo, and KAGRA collaborations \citep{Abbott2016,Abbott2023}, have further motivated the Laser Interferometer Space Antenna (LISA) mission \citep{LISA_white_paper, lisa_multimessenger_whitepaper,lisa_gw_whitepaper,lisa_consortium_waveform_working_group_waveform_2023,Amaro-Seoane2023}, which will enable multimessenger astronomy with gravitational waves of supermassive binary black holes. None of the binary black hole GW signals detected so far has been accompanied by an unequivocal electromagnetic (EM) counterpart, due to the "dry" environment expected around stellar-mass binary black holes~\citep{Lynden-Bell1969,Abbott2023}. However, an SMBBH could be surrounded by a hot accretion flow in a fashion similar to regular AGNs, making these sources highly promising multi-messenger sources with GWs. Thus, a major goal of studies of accretion onto SMBBHs is to identify EM counterparts to the gravitational wave emission expected during the final stages of black hole coalescence.

The study of accretion onto SMBBHs can also help understand how the interaction between the disk and the binary influences the binary's orbital evolution. Angular momentum loss to the stars occurs readily at large radii but slows down at parsec-scale separations \citep{Begelman1980,Milosavljevi2001}. Before binary-disk decoupling, the interaction between the disk and the binary can help to exchange angular momentum between the disk and the secondary, reducing the binary orbital separation. The tidal torques of the binary tend to drive the gas away from the orbit, clearing a lower density "gap" around the binary. Viscous torques transport the angular momentum outward in the disk, allowing gas to flow inward and refill the cavity. As a result, the disk can be tidally truncated by the secondary, further reducing the separation \citep{Lin+79, Lin1986, Armitage2002,Hayasaki2008,Ivanov1999,Farris2014}. 

A reliable description of accretion flows and their associated radiation in merging SMBBH requires a radiation magnetohydrodynamics simulation in full general relativity. However, computational limitations make such simulations currently infeasible. Newtonian hydrodynamic simulations incorporating some of the relevant physics have been performed at various levels of approximation, see, e.g., \citep{Rossi2010,Corrales2010,Orazio2013MNRAS.436.2997D,Neill2009,Diego2019,Duffell2020,westernacher-schneider_multi-band_2022,Clyburn:2024xyq,delaurentiis_relativistic_2024,Tiwari:2025imm} and references therein. General relativistic simulations are underway see \citep{gold_relativistic_2019,Cattorini:2023akr} for reviews, and~\cite{Ruiz:2023hit,Manikantan:2024giq,Avara:2023ztw,Ennoggi:2025nht,Manikantan:2025afy,Ressler:2024tan} and references therein for more recent work. Circumbinary accretion can be affected by many parameters, including the accretion rate, disk inclination, equation of state, viscosity, and magnetic field. Additionally, binary separation, mass ratio, orbital eccentricity, and black hole spins also play significant role \citep{Duffell2020,Paschalidis:2021ntt,Bright:2022hnl,Dittmann2022,Wang2023,Turpin2024,Manikantan:2025afy}. Finally, orbit-averaged, stationary models have been helpful in identifying salient bulk features of circumbinary disks and their emission. However, these have relied on the standard thin disk models to describe internal stresses, see, e.g., \citep{Milosavljevi2001,Liu+10,Tanaka2010,Kocsis:2012cs}.

In this paper, we focus on orbit-averaged models but go beyond the thin disk model, and develop a stationary, geometrically thick slim disk model of accretion  onto  binary black holes. Slim disk models are better suited than thin disk ones when the accretion rate is approaching or exceeds the Eddington limit~\citep{Abramowicz1988}. Our model incorporates the advective cooling effect, and allows the gas to deviate from circular Keplerian motion~\citep{Abramowicz1988}. Our model is applicable to SMBBH disks 
well before binary-disk decoupling, with accretion rates ranging up to super-Eddington values and with low mass ratios. We adopt the Paczy{\'n}sky-Wiita pseudo-Newtonian gravitational potential \citep{Paczy1980} for the primary black hole, and treat the secondary's potential as a perturbation. The influence of the secondary is considered in the vertical, radial, and azimuthal directions. We numerically solve the cicrumbinary slim disk  equations, and compare the results with a thin disk model. We investigate how different parameters influence the disk solutions and quantify the disk's efficiency in driving the binary inward migration.

This paper is organized as follows. We present the equations of our model in Section \ref{sec:level1}. We compare the thin disk results with the slim disk results in Section \ref{thinvsslim}. In Section \ref{results}, we show the details of slim disk solutions. We conclude with a discussion of our model in Section \ref{conclution}.

\section{\label{sec:level1}Stationary Slim disk model}

In this section, we derive the equations for both slim and thin circumbinary disks. Additionally, we calculate the disk spectrum based on the assumption of multi-color black-body radiation.

\subsection{The slim disk equations}
We use cylindrical coordinates to describe the disk equations. The equations governing axisymmetric accretion flow in the non-relativistic limit are given by the Navier-Stokes equations:
\begin{eqnarray}
\label{N-S1}
&&\frac{ \partial \rho}{\partial t}+ \bigtriangledown (\rho \vec{u})=0,\\
\label{N-S2}
&&\rho\left(\frac{ \partial \vec{u}}{\partial t}+(\vec{u} \cdot \bigtriangledown) \vec{u}\right)=-\rho \bigtriangledown \Phi - \bigtriangledown p+ \bigtriangledown T_{\rm vis}.
\end{eqnarray}
Here $\rho$, $p$, $\vec{u}$ and $T_{vis}$ are gas density, gas pressure, gas velocity and gas viscous stress tensor, respectively. $\Phi=U_{\rm p}+U_{\rm s}$ is the total gravitational potential, that has contributions from both the primary and the secondary black hole. We consider a pseudo-Newtonian potential \citep{Paczy1980} for the primary
\begin{equation}
U_{\rm p}=-\frac{GM_{\rm p}}{R-2R_g}, ~~(R^2=r^2+z^2, R_g=GM_{\rm p}/c^2),
\end{equation}
where $G$ is the gravitational constant, $M_{\rm p}$ is the primary's mass, $r$ the cylindrical radius, and $c$ is the speed of light. The gravitational potential due to the secondary $U_{\rm s}$ is small and is given by
\citep{Lin+79,Peter+80},
\begin{equation}
U_{\rm s}=\frac{-Gm_{\rm s}}{\sqrt{a^2+R^2-2ar\cos(\phi-\omega t)}}+\frac{Gm_{\rm s}r\cos(\phi-\omega t)}{a^2}.
\end{equation}
Here, $\omega$, $m_{\rm s}$ and $a$ are the angular velocity of the secondary, the mass of the secondary, and the binary separation, respectively. The second term arises because $r$ is measured from the primary, and hence the frame of reference is not inertial. In this work, we focus on a steady state disk. The orbit-averaged derivatives of $U_{\rm s}$ with respect to r, $\phi$ and z are:
\begin{eqnarray}
\label{conservation of mass}
&&U_{\rm s,r}=\frac{1}{2\pi}\int_{0}^{2\pi}\frac{\partial U_{\rm s}}{\partial r}\,d\theta = \frac{1}{2\pi}\frac{\partial}{\partial r}\int_{0}^{2\pi} U_{\rm s}\,d\theta,\\
&&U_{\rm s,\phi}=\frac{1}{2\pi}\int_{0}^{2\pi}\frac{\partial U_{\rm s}}{\partial \phi}\,d\theta=-\Lambda,\\
&&U_{\rm s,z}=\frac{1}{2\pi}\int_{0}^{2\pi}\frac{\partial U_{\rm s}}{\partial R}\frac{z}{R}\,d\theta=\frac{1}{2\pi}\frac{z}{R}\frac{\partial}{\partial R}\int_{0}^{2\pi} U_{\rm s}\,d\theta,
\end{eqnarray}
where $\theta=\phi-\omega t$. $\Lambda$ is the specific tidal torque, describing the rate of change of specific angular momentum of the disk due to the binary tidal torque.
%It can be approximates as $\Lambda=f\rm{sign}(r-a)m_{\rm s}^2r/(\Omega^2(r-a)^4)$ \citep{Lin+79,Goldreich1980,Armitage2002,Chang2008,Lodato+09,Liu+10}, where f is a dimensionless normalization factor ($f\sim0.01$ \citep{Armitage2002}), $a$ is the binary separation, $q$ is the mass ratio. 
Following \cite{Lodato+09}, we use the commonly adopted approximation for $\Lambda$~\citep{Armitage2002MNRAS, Armitage2002,Chang2008,Liu+10},
\begin{equation}
\label{Lambda}
\Lambda=\rm{sign}(r-a)\frac{fq^2 M_{\rm p}}{2r}\left(\frac{\rm{min}(r,a)}{\Delta_{\rm p}}\right)^4,
\end{equation}
where $q=m_s/M_p$ is the mass ratio, $\Delta_{\rm p}=r-a$, and $f=0.01$ is a dimensionless normalization factor \citep{Armitage2002}. 
We smooth the torque when $r\approx a$ by setting $\Delta_{\rm p}= {\rm max}(|r-a|,R_{\rm H})$, where $R_H=(q/3)^{1/3}a$ is the Hill sphere of the secondary.

When orbit averaging $U_s$ we have integrals of the form
\begin{equation}
\frac{1}{2\pi}\int_{0}^{2\pi}\frac{1}{\sqrt{1+r_a^2+z_a^2-2r_a \cos(\theta)}}\,d\theta,
\end{equation}
where $r_a=r/a$ and $z_a=z/a$. To obtain a closed form for these integrals that facilitate the integration of the disk structure equations, we exploit the fact that the main contribution to the integral comes from $\theta\approx 0$, and approximate $\cos\theta$ using $\cos\theta\approx1-\theta^2/2$. The error introduced by this approximation for $r_a>0$ is $<13\%$, and makes the numerical integration of the disk structure equations feasible. With this approximation, we first obtain the orbit averaged $U_{\rm s}$ as, 
\begin{eqnarray}
\nonumber
\bar U_s&\approx&\frac{-Gm_s}{a\pi\sqrt{r_a}}\ln \left( \frac{\sqrt{r_a} \pi + \sqrt{(1 - r_a)^2 + z_a^2 + r_a \pi^2}}{\sqrt{(1 - r_a)^2 + z_a^2}} \right)   \\
&\approx&\frac{Gm_s}{a\pi\sqrt{r_a}}\ln \left(\frac{\sqrt{(1 - r_a)^2 + z_a^2}}{2\pi\sqrt{r_a}}\right).\label{eq:barUsapprox}
\end{eqnarray}

The second equation holds, because $\frac{r_a\pi^2}{(1-r_a)^2+z_a^2}\gg 1$ when $r_a\sim 1$. The error introduced by the second approximation in Eq.~\eqref{eq:barUsapprox} is less than $13\%$, 
as verified by numerical integration of $U_s$. 
The two orbit-averaged derivatives of $U_{\rm s}$ in the radial and vertical directions can be calculated as,
\begin{widetext}
\begin{eqnarray}
&&U_{s,r}=\frac{\partial \bar U_s}{\partial r_a}\frac{\partial r_a}{\partial r}=\frac{Gm_s}{a^2\pi \sqrt{r_a}}\left(\frac{r_a-1}{(r_a-1)^2+z_a^2}-\frac{1}{2r_a}\right)-\frac{\bar U_s}{2ar_a}, \\
&&U_{s,z}=\frac{\partial \bar U_s}{\partial z_a}\frac{\partial z_a}{\partial z}=\frac{Gm_s}{a^2\pi \sqrt{r_a}}\frac{z_a}{(r_a-1)^2+z_a^2}.
\end{eqnarray}
\end{widetext}

When $r_a=1$, both equations diverge. Following the treatment of tidal torque $\Lambda$, when $|r-a|<R_H$, we set $|1-r_a|=R_H/a$.

We now discuss the stationary slim disk equations.

(i) The mass conservation:

From Eq. \eqref{N-S1}, assuming $u_z=0$, $\frac{\partial (\rho u_\phi)}{\partial \phi}=0$ and integrating the equation vertically between the disk's vertical boundaries at $z = \pm H$, we obtain
\begin{equation}
\frac{\partial \Sigma}{\partial t}+ \frac{1}{r}\frac{\partial (r \Sigma V)}{\partial r}=0,
\end{equation}
here,
\begin{eqnarray}
&&\Sigma=\int_{-H}^{+H}\rho\,dz,~V=\frac{1}{\Sigma}\int_{-H}^{+H}\rho u_r\,dz.
\end{eqnarray}
For a stationary disk, the mass conservation equation is given by:
\begin{equation}
 \dot M=-2\pi r \Sigma V.
\label{sigma}
\end{equation}
Here, $\dot M$ is the rest-mass accretion rate and is an integration constant.

(ii) The vertical hydrostatic equilibrium:

Taking the z direction in Eq.~\eqref{N-S2} and orbit averaging, we obtain
\begin{eqnarray}
\label{H2}
&& \frac{\partial p}{\partial z}=-\rho (\Omega_{\rm k}^2 z+ U_{\rm s,z}),
\end{eqnarray}
where $\Omega_{\rm k}=(1/r)(\partial U_p/\partial r)=(GM_{\rm p})^{1/2}r^{-3/2}(1-2R_g/r)^{-1}$. 

Following  \cite{sadowski2011}, we assume  $p \propto \rho^{1+1/N}$. When solving the equations numerically, for simplicity we set $N = 3$, which corresponds to radiation pressure domination. Integrating this equation vertically between the center disk and the vertical boundary at $z = H$, we obtain:
\begin{equation}
\label{H3}
\frac{B_iP}{\Sigma}= (\Omega_{\rm k}^2+ \Omega_{\rm s}^2)H^2,
\end{equation}
where
\begin{equation}
B_i=2N+2, P=\int_{-h}^{+h}p\,dz, ~\Omega_{\rm s}^2=2U_{\rm s,z}(z=H/2)/H.
\end{equation}

To derive Eq. \ref{H3}, we assumed $U_{\rm s,z}=U_{\rm s,z}(z=H/2)$. This approximation introduces errors of less than $20\%$ when compared to the properly z-integrated gravitational force contribution from the secondary. However, our code cannot converge to solve the structure equations when we adopt the full analytic expression, as the equations become very complex. This approximation allows us to integrate the disk structure equations. The error introduced by this approximation has negligible impact on our solutions, because, as we discuss later, the gravitational force contribution in the z direction from the secondary is subdominant. 

Note that, different treatment on the integration result in different $B_i$, for example the original slim disk treatment yields $B_i=6$ \citep{Abramowicz1988}, the GR slim disk treatment gives $B_i=2$ \citep{Abramowicz1996}, and the approximation $\frac{\partial p}{\partial z}\sim-p/z$ leads to $B_i=1$. According to \cite{sadowski2011}, the polytropic assumption yields a flux profile that is approximately the same as a model with full vertical hydrostatic equilibrium (a "two-dimensional" model), despite that it has a substantial impact on the disk thickness and density profile. We tested different choices for $B_i$, and find that our main conclusions are not qualitatively sensitive to $B_i$; see  Appendix for further details.

(iii) The angular momentum conservation:

Taking into account the $\phi$ direction in Eq. \eqref{N-S2}, we obtain,
\begin{equation}
\rho ru_r \frac{\partial u_\phi}{\partial r}+\rho u_r u_\phi= \frac{1}{r} \frac{\partial (r^2T_{r\phi})}{\partial r}-\rho \frac{\partial U_{\rm s}}{\partial \phi}.
\end{equation}
Adopting the standard $\alpha$ disk recipe, $T_{r\phi}=-\alpha P$, and integrating vertically we obtain the angular momentum conservation equation,
\begin{equation}
\label{angular}
-\Sigma V\frac{\partial (\Omega r^2)}{\partial r}=\alpha r\frac{\partial P}{\partial r}+2\alpha P-\Sigma \Lambda.
\end{equation}
To derive the last equation we assumed that $\Lambda$ has negligible depenendence on $z$ as is common in the literature.

(iv) The radial momentum conservation:

The r direction of Eq.\eqref{N-S2} yields
\begin{equation}
V\frac{ \partial V}{\partial r}+(\Omega_{\rm k}^2-\Omega^2)r+ \frac{1}{\Sigma}\frac{\partial P}{\partial r}=- U_{\rm s,r}(z=H/\sqrt{5}).
\label{V_r}
\end{equation}
To derive this equation, we assumed that $\int^H_{-H}\rho U_{\rm s,r}dz=\Sigma U_{\rm s,r}(z=H/\sqrt{5})$. 
This choice is motivated by the fact that both $\rho$ and $U_{\rm s,r}$ must become maximum at $z=0$ and as such, they must have a quadratic dependence on $z$ for $z/r\ll 1$ and $z/H\ll 1$. Expanding $U_{\rm s,r}$ in powers of $z$ and assuming $r_a\gg 1$ we obtain $U_{s,r}=U_{s,r}(z=0)[1-(z/r)^2]$. The density must scale as $\rho=\rho_0[1-(z/H)^2]$, where $H$ is $r$-dependent. Under these scalings, it can be shown that $\int^H_{-H}\rho U_{\rm s,r}dz=\Sigma U_{\rm s,r}(z=H/\sqrt{5})$. \ 
Our tests show that $U_{s,r}$ has negligible
impact on the solutions (see Sec.~\ref{Sec:impact_of_Usr_Usz} below), thus the aforementioned approximation does not affect our main conclusions.

% \sw{I get $\sqrt{1/5}$. setting $\rho=\rho_0(1-(Z/H)^2)$, $U_{s,r}=U_{s,r,0}(1-Z^2/(r_a-1)^2)$(after expanding the most important term in the equation)}  \sw{Oh, It is $z_a$ not $Z$ I made a mistake.  The important term in $U_{s,r}$ is $\frac{Gm_s}{a^2\pi \sqrt{r_a}}\left(\frac{r_a-1}{(r_a-1)^2+z_a^2}-\frac{1}{2r_a}\right)$, and only $\frac{r_a-1}{(r_a-1)^2+z_a^2}$ has $z_a$. It can be written into $\frac{Gm_s}{a^2\pi \sqrt{r_a}r_a-1}(1+(\frac{z_a}{r_a-1})^2)^{-1}$. I use $(1+x)^{-1}\approx1-x$ to expend the $U_{s,r}$}. 

(v) Thermal balance equations:

The advected heat  $Q^{\rm adv}$ at each radius is equal to the difference between viscous-tidal heating and radiative cooling:
\begin{equation}
\label{Qadv}
Q^{\rm adv}=Q^{\rm vis}+Q^{\rm tid}-Q^{\rm rad}.
\end{equation}

Employing the diffusive approximation and substituting differentials with finite differences \citep{sadowski2011}, we obtain: 
\begin{equation}
\label{Qrad}
Q^{\rm rad}(r,z)=-\frac{16\sigma T^3}{3\kappa \rho}\frac{\partial T}{\partial z},~~~ Q^{\rm rad}=\frac{32\sigma T^4}{3\Sigma\kappa}.
\end{equation}
The opacity coefficient $\kappa$ is calculated by using the Kramer's approximation:
\begin{equation}
\label{e.kramers}
\kappa=\kappa_{\rm es}+\kappa_{\rm ff}=0.34+3.2\times 10^{22} T^{-3.5}\Sigma/H.
\end{equation}

The viscous heating $Q^{\rm vis}$ can be calculated by \citep{Pringle1981,Abramowicz1988}:
\begin{equation}
\label{Qvis}
Q^{\rm vis}=-\frac{\alpha Pr}{2}\frac{d\Omega}{dr}.
\end{equation}
For simplicity, we set $d\Omega/dr=d\Omega_k/dr$ \footnote{We also solve the equation using $d\Omega/dr$. The results are similar to those obtained using $d\Omega_k/dr$ since the $\Omega$ of the slim disk only slightly deviates 
from the Keplerian values, resulting in a small change in $Q^{\rm vis}$. However, using $d\Omega_k/dr$ prevents the solution from having a negative $Q^{\rm vis}$ near $r=a$.}.

The tidal torque heating $Q^{\rm tid}$ can be calculated as \citep{Lodato+09}:

\begin{equation}
\label{Qtid}
Q^{\rm tid}=\frac{1}{2}(\Omega(a)-\Omega(r))\Lambda\Sigma.
\end{equation}
Since $Q^{tid}$ becomes important near $a$, we perform a Taylor expansion to calculate the term  $\Omega(a)-\Omega(r)=\frac{d\Omega_k}{dr}(a-r)$, and when $|a-r|<R_H$, we set $|a-r|=R_H$. This procedure is the same as the process adopted to calculate the $\Lambda$. 

We now calculate the advective cooling ($Q^{\rm adv}$) from the equation of state and the first law of thermodynamics.
The total pressure is the sum of gas pressure $p_{\rm g}=k_B\rho T/(\mu m_{\rm p})$ and radiation pressure $p_{\rm r}= a_cT^4/3$, where $T$ is the gas temperature, $k_B$, $m_{\rm p}$ and  $a_c$ are the Boltzmann constant, the proton mass and the radiation constant, respectively. The mean molecular weight is taken to be $\mu=0.62$, for a fully ionized gas with solar abundance \citep{Cox1968}. 
The equation of state can be expressed in the form
\begin{equation}
\label{EOF}
P=\frac{k}{\mu m_{\rm p}}\Sigma T+\frac23Ha_cT^4.
\end{equation}

We write the specific internal energy as 
\begin{equation}
\label{u}
U=\frac{3p_{\rm r}}{\rho}+\frac{3p_{\rm g}}{2\rho}.
\end{equation}
The first law of thermodynamics can be written as,
\begin{equation}
\label{du}
dU=TdS-pd\rho^{-1},
\end{equation}
 where $S$ is the specific entropy. 
 %From Eq. \ref{u} and Eq. \ref{du}, we get
% \begin{equation}
% \frac{\Sigma T}{P}\frac{dS}{dr}=3\frac{d \ln P}{dr} -\frac{3\beta}{2}\frac{d \ln T}{dr}-4\frac{d \ln \Sigma}{dr}.
% \end{equation}
The advective heating is written as  \citep{Abramowicz1988},
\begin{equation}
Q^{\rm \rm adv}=\frac{V\Sigma T}{2}\frac{dS}{dr}.
\end{equation}
Here, $Q^{adv}$ has been divided by 2, so that it is consistent with $Q^{\rm tid}$, $Q^{\rm vis}$ and $Q^{\rm rad}$, which account for only one side flux.

\subsection{Numerical solution}
There are a total of 6 independent equations:~Eq.\eqref{sigma}, Eq.\eqref{H3}, Eq.\eqref{angular}, Eq.\eqref{V_r}, Eq.\eqref{Qadv} and Eq.\eqref{EOF}. 
These equations involve six variables: $\Sigma$, P, H, T, V and $\Omega$. Using Eq.\eqref{sigma}, Eq.\eqref{H3}, and Eq.\eqref{EOF}, $\Sigma$, P, and H can be eliminated as independent variables, leaving 3 independent variables: T, V and $\Omega$. These variables are determined by solving three ordinary differential equations: Eq.\eqref{angular}, Eq.\eqref{V_r} and Eq.\eqref{Qadv}.
After some manipulations, Eq.\eqref{V_r} and Eq.\eqref{Qadv} can be written as,
\begin{eqnarray}
&& a_1 \frac{d \ln V}{dr}+b_1\frac{d \ln T}{dr}=c_1,\\
&& a_2 \frac{d \ln V}{dr}+b_2\frac{d \ln T}{dr}=c_2,
\end{eqnarray}
where,
\begin{eqnarray}
\nonumber
&&a_1=\frac{V^2}{c_s^2}+\frac{1-3\beta}{\beta+1},~b_1=\frac{8-6\beta}{\beta+1},\\\nonumber
&&c_1=\frac{(\Omega^2-\Omega^2_k)r-U_{\rm s,r}}{c_s^2}-P_3,\\\nonumber
&&a_2=\frac{3-9\beta}{\beta+1}+4,
%&&a_2=\left(3-\frac{ \alpha^2 c_s^2}{V^2}\right)\frac{1-3\beta}{\beta+1}+4,\\\nonumber
b_2=\frac{24-18\beta}{\beta+1}-\frac{3\beta}{2},\\ \nonumber
%&&b_2=\left(3-\frac{ \alpha^2 c_s^2}{V^2}\right)\frac{8-6\beta}{\beta+1}-\frac{3\beta}{2},\\ \nonumber
&&c_2=\frac{2}{VP}\left(Q^{\rm tid}+Q^{\rm vis}-Q^{\rm rad}\right)-s_3, \\\nonumber
%&&c_2=\frac{2}{VP}\left(Q^{\rm tid}-Q^{\rm rad}\right)+\frac{\alpha}{rV^2}\left(2\alpha c_s^2-\Lambda\right) \\ \nonumber
%&&+\frac{2\alpha\Omega}{V}-s_3+\frac{ \alpha^2 c_s^2}{V^2}P_3, \\\nonumber
&&P_3=\frac{1-3\beta}{(\beta+1)r}
-\frac{1-\beta}{\beta+1}\frac{d \ln(\Omega_k^2+\Omega_{\rm s}^2)}{dr}, \\\nonumber
&&s_3=3P_3+\frac{4}{r},~c_s^2=P/\Sigma.
\end{eqnarray}
%here, $c_s^2=P/\Sigma$.
Here $\beta=p_{\rm g}/p$. Thus the three ODEs that determine the system can be rewritten as
\begin{eqnarray}
\label{10}
&&\frac{d \ln V}{d r}=\frac{\mathcal{N}}{ \mathcal{D}}=\frac{c_1b_2-c_2b_1}{a_1b_2-a_2b_1}, \\
\label{11}
&&\frac{d \ln T}{d r}=\frac{c_2}{b_2}-\frac{a_2}{b_2}\frac{\mathcal{N}}{\mathcal{D}},\\
\label{12}
&&\frac{d L}{d r}=-\frac{1}{V}\left(\alpha rc_s^2\frac{d \ln P}{d r}+2\alpha c_s^2-\Lambda\right).
\end{eqnarray}

These equations require a boundary condition.
Taking $T_{\phi r}=0$ at the inner boundary and integrating the equation of angular momentum conservation, we obtain
\begin{equation}
\frac{\dot M}{2\pi}(L-L_i)=-r^2T_{\phi r}- \int_{r_i}^{r}r\Sigma\Lambda\,dr,
\label{L_i}
\end{equation}
here, $L_i$ is the specific angular momentum of the debris in the inner boundary. The term $\int_{r_i}^{r}r\Sigma\Lambda\,dr$ in Eq.~\eqref{L_i} is the integrated tidal torque, with $r_i$ the radius at the disk inner edge. We set,
\begin{equation}
L_0=L_i- \frac{2\pi}{\dot M}\int_{r_i}^{r}r\Sigma\Lambda\,dr.
\label{L_0}
\end{equation}
$L_0$ is the difference between the specific angular momentum of gas at the inner boundary and the integrated specific angular momentum imparted on the gas by the tidal torque. At the outer edge ($R_{\rm out}=2\times 10^4 R_g$), as $T_{\rm tid}$ and $Q^{\rm adv}$ are small, we assume the disk reduces to the Shakura-Sunyaev disk, $L\approx L_{\rm k}=\Omega_{\rm k}r^2$. From Eqs. \eqref{L_i} and \eqref{L_0}, we obtain,
\begin{equation}
  P=\frac{(L_{\rm k}-L_0)\dot M}{2\pi \alpha r^2}.  
\end{equation}
Taking $Q^{\rm rad}= Q^{\rm vis}$ at the outer edge, we obtain
\begin{equation}
T^4=\frac{9\alpha \kappa M_{\rm p}^{0.5}\Sigma P}{128 \sigma r^{1.5}}. 
\end{equation}
Here, we define the Eddington accretion rate as,
\begin{equation}
    \dot M_{\rm Edd}=1.37 \times 10^{23} ~{\rm kg~s}^{-1} \eta^{-1}_{-1} M_8,
\end{equation}
where, $\eta$ is the radiative efficiency of accretion, $\eta_{-1}=\eta/0.1$, $M_8=M_{\rm p}/(10^8M_\odot)$, and $M_\odot$ is the solar mass.
Hereafter, we refer to accretion rates in dimensionless Eddington units,
i.e., $\dot m= \dot M/ \dot M_{\rm Edd}$.

The free parameters are $M_{\rm p}$, $\dot m$, $\alpha$, $a$ and mass ratio $q$. Given $L_{\rm k}$ and $L_0$, one can construct the boundary value of $T_0$, $V_0$ and $L$ at the outer edge. 
$L_0$ has to be chosen properly to yield $\mathcal{N}=\mathcal{D}=0$ at the sonic point $r_s$ \citep{Abramowicz1988}.  We use the 4th-order Runge-Kutta method to integrate the differential equations, employing a shooting technique to search for the sonic point. If ${L}_{0}$ is greater than the true value, then $\mathcal{D}$ decreases to 0 before it reaches the real sonic point (the solution breaks down before it reaches the sonic point). On the other hand, if ${L}_{0}$ is smaller than the true value, then $\mathcal{D}$ will not decrease to 0 (the equations can be solved down to the event horizon of the primary, but yield a divergent $T$). We iterate to locate the appropriate value of ${ L}_{0}$, until $|\Delta {L}_{0}/{L}_{0}|$ is less than $10^{-7}$, with $\Delta {L}_{0}$ the difference in the two consecutive values of $L_0$ in our iterative scheme. We note that our solutions are insensitive to the specific choice of $T_0$ and $V_0$.

\subsection{The disk thermal spectrum}
Assuming blackbody radiation at each annulus of radius $r$, the specific radiative intensity can be written as,
\begin{equation}
\label{Iv}
 I_{\rm d}(\nu,r)=\frac{2h\nu^3c^{-2}}{\exp(h\nu/k_{\rm B} T_{\rm e}(r))-1}.
\end{equation}
Here, $h$ is the Plank constant, $k_{\rm B}$ is the Boltzmann constant, and $T_e$ is the effective temperature $T_e=(Q^{\rm rad}/2\sigma)^{1/4}$ ($\sigma$ is Stefan-Boltzmann constant). The local spectrum can be calculated as, 
\begin{equation}
F(\nu)=\frac{1}{d^2}\int 2\pi rI_{\rm d}(\nu,r)\,{\rm d}r,
\label{flux_o}
\end{equation}
where $d$ is the luminosity distance from the observer to the disk. By integrating the spectrum from $R_{\rm out}$ to the innermost stable circular orbit (ISCO), one can calculate the spectra for different disk parameters.

\section{Comparing with the thin disk solution} \label{thinvsslim}

In this section, we reproduce the stationary binary thin disk solution of \cite{Liu+10}, and compare its result with the slim disk results. In \cite{Liu+10}, the viscous torque is calculated as $T_{\phi r}=-3\nu\Sigma \Omega_{\rm k}/2$, where $\nu(r)=\nu_0r^{\lambda}$. The angular momentum conservation equation is written as,
\begin{equation}
\frac{\partial \ln \Sigma}{\partial r}=\left(\frac{\dot M }{3\nu_0\pi r^{\lambda}\Sigma}-1\right)\frac{\partial \ln L_k}{\partial r}+\frac{2\Lambda r^{1-\lambda} }{3\nu_0L_k}-\frac{\lambda}{r}.
\label{thind}
\end{equation}
Here, we set $\lambda=1$ as in~\cite{Liu+10}. An alternative way to solve for the thin disk model is to consider $T_{\phi r}=-\alpha P$. By using Eq. \eqref{H2}, ignoring the term $\Omega_s$ and treating $H/r$ as a constant as in \cite{Liu+10}, the angular momentum conservation equation takes the following form,
\begin{equation}
\frac{\partial \ln \Sigma}{\partial r}=\left(\frac{\dot M L_k}{2\pi \alpha r^2 P}-2\right)\frac{\partial \ln L_k}{\partial r}+\frac{\Lambda \Sigma }{\alpha r P}.
\label{thinP}
\end{equation}

Following \cite{Liu+10}, we impose the inner boundary condition at the ISCO, where the viscous torque $T_{\phi r}$ decreases to zero. Then, the angular momentum conservation equation can be solved for the given $M_{\rm p}$, $\dot M$, $\nu_0$ (or $\alpha$) and $\Sigma_{\rm out}$ at the disk outer radius. Due to the inner boundary condition, only two of the three parameters-- $\dot M$, $\Sigma_{\rm out}$ and $\nu_0$ (or $\alpha$)--remain independent. We adopt the 4th-order Runge-Kutta method to integrate the Eqs.~\eqref{thind} and~\eqref{thinP}, and employ the shooting technique to determine the parameter $\xi$ (one of $\dot M$, $\Sigma_{\rm out}$ and $\nu_0$). If $\xi$ is greater than the true value, then the solution breaks down before it reaches the ISCO, while if $ \xi$ is smaller than the true value, $\Sigma$ will not decrease to 0 when $r$ approaches the ISCO. We iteratively search for the true $\xi$, until $|\Delta  \xi/ \xi|$ is less than $10^{-7}$.

\cite{Liu+10} use a integration method to solve the equation, given the initial value of $\nu_0$ and $\Sigma_0$ in the outer radius of the disk $R_{\rm out}$. $\dot M$ is then determined by the inner boundary condition $T_{\phi r}(r_{\rm ISCO})=0$. 
Our code reproduces the results from table I of \cite{Liu+10} in most cases. However, for some cases, $\dot{m}$ is sensitive to the input parameters, and the number of significant figures in Table~I of \cite{Liu+10} is unclear. Nevertheless, even in those cases, our $\dot{m}$ agrees with that of \cite{Liu+10} to within $10$--$20\%$. When comparing the thin disk solutions with the slim disk solutions, we set the $\dot m$ and $\Sigma_{\rm out}$ at the same value as those for slim disk, and determine the $\nu_0$ from  the inner boundary condition. We set $R_{\rm out}=2\times10^4R_g$ for both the slim disk and the thin disk.
\begin{figure}
    \centering    
    \includegraphics[width=0.46\textwidth]{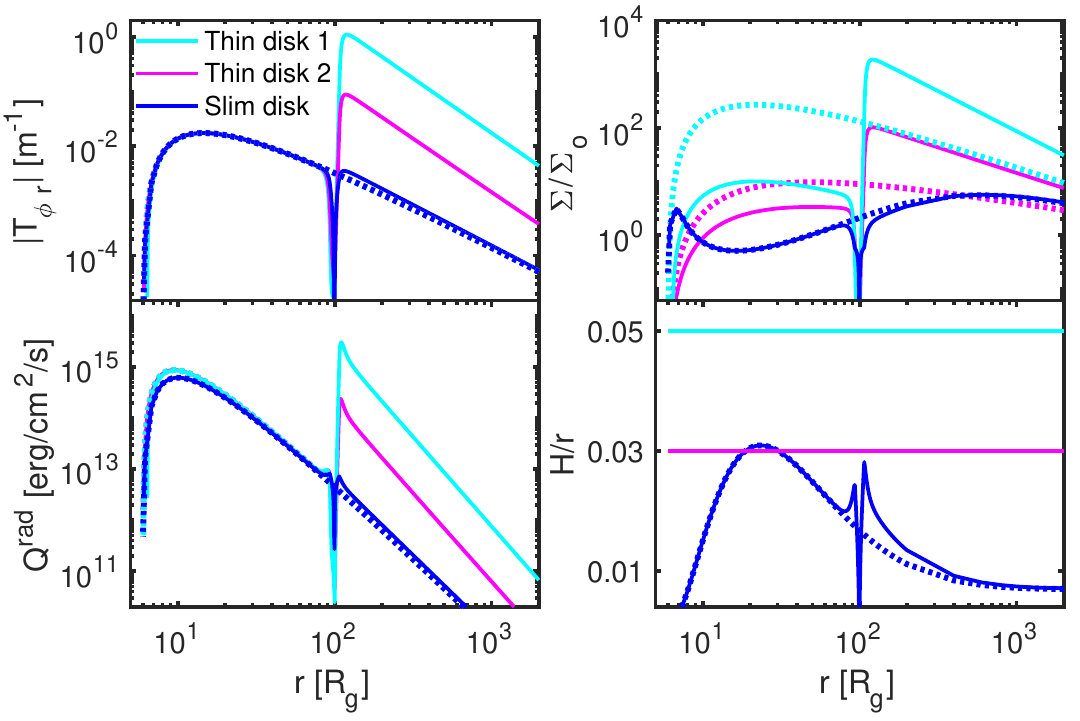}   %\includegraphics[width=0.46\textwidth]{qt_vs_qv.pdf} 
    \includegraphics[width=0.46\textwidth]{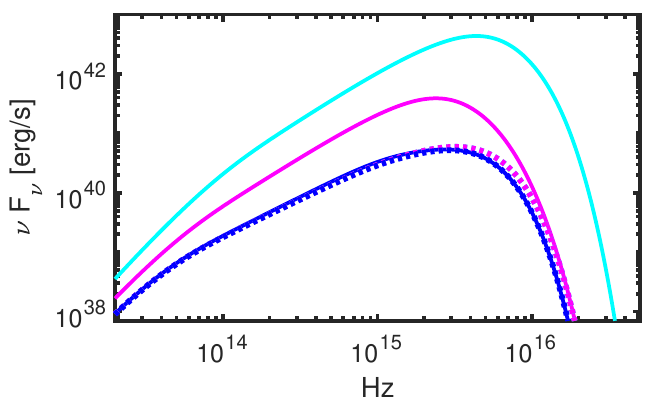} 
    \caption{%\vp{I recommend we remove the plot on the ratio of the tidal to viscous heating, unless it can become smoother in the thin disk cases .}
    Slim disk results VS. thin disk results. Top panel (left to right column): viscous torque $T_{\phi r}$, radiative flux $Q^{\rm rad}$, surface density, and $H/r$ profiles for the different disk models. Magenta lines and cyan lines denote the thin disk results for cases of $T_{\phi r}=-3\nu \Sigma\Omega_k/2$ and $T_{\phi r}=-\alpha P$, respectively, while the blue lines show the slim disk results. The color dotted lines denote $q=0$, while the color solid lines represent $q=10^{-3}$ for cases of both slim and thin disks. The disk parameters are $\dot m=0.1$, $M_{\rm p}= 10^8M_\odot$, $a=100R_g$ for all the three models.}
    \label{fig:thin_vs_slim}
\end{figure}

Figure \ref{fig:thin_vs_slim} illustrates the disk solutions and spectra for both slim and thin disks. For the slim disk cases, we set $\alpha=0.1$. When $q=0$ (dotted lines), both slim and thin disks exhibit similar values of $T_{\phi r}$. This similarity arises from the expression $T_{\rm \phi r}=-\dot{M} (L-L_i)/(2\pi r^2)$ (see Eq. \eqref{L_i}), which holds for both slim and thin disks when $q=0$. Although slim disks assume $T_{\phi r}=0$ near the event horizon, they confirm the termination of the disk at the ISCO for low accretion rates, aligning with the assumption of thin disks. The slim disk solution indicates that when the accretion rate is low, the disk terminates at the ISCO, and the angular momentum can be approximated by $L_k$. Consequently, the solution for $T_{\phi r}$ resembles that of a thin disk. For this reason, the spectrum of a slim disk would reduce to that of a thin disk for low accretion cases (see the dotted color lines in the bottom plot of Figure \ref{fig:thin_vs_slim}). It is important to note that the solution of $T_{\phi r}$ is insensitive to the specific prescription of $T_{\phi r}$, as depicted in the figure. Despite the similarity in $T_{\phi r}$ solutions between slim and thin disks, their density profiles differ significantly, as illustrated in the upper right panel. This discrepancy arises because the inflow velocity and disk thickness (see lower right panel on the upper plot of Figure \ref{fig:thin_vs_slim}) both influence the density profile, as shown in Eq. \eqref{sigma} and Eq. \eqref{H3}. Consequently, any perturbation in the density profile leads to substantial differences between slim and thin disks.

For cases where $q=10^{-3}$ (solid lines), $T_{\phi r}$ experiences a sharp drop near  $r=a$ while remaining nearly unchanged for the inner part of the disk for both slim and thin disk solutions when comparing with case of $q=0$. As shown in the figure, at $r > a$, $T_{\phi r}$ is increased by a factor of one to two orders of magnitudes in the thin disk cases, while it is only slightly higher than the case of $q=0$ for slim disk. At $r > a$, $T_{\phi r}$ receives an additional contribution due to the secondary, given by $\int_{r_i}^{r} r \Sigma \Lambda dr$. Conversely, at $r < a$, the tidal torque becomes significantly smaller, and the tidal torque onto% \vp{"onto" instead of "from" since the torque comes from the binary?} 
the outer disk does not contribute to the inner viscous torque. These factors result in the inner viscous torque being minimally affected by the secondary. The presence of the secondary does not alter substantially the thermal emission of the inner part of the disk,  but can induce a significant boost in viscous heating at $r > a$, and increase the disk's brightness near \( r = a \) because of tidal heating. 

The integrated tidal torque strongly depends on the disk density profile.  As discussed earlier, different viscosity prescriptions significantly affect the disk surface density, thereby influencing $T_{\phi r}$. As observed in the figure, a thin disk yields a stronger $T_{\phi r}$ than a slim disk at $r > a$. This is because a thin disk has a higher surface density compared to the slim disk. The disk density profile is boosted by orders of magnitude when compared to the case of $q=0$ for thin disks, while it becomes even smaller for slim disks at $r > a$. For this reason, the integrated tidal torque is significantly smaller for slim disks than for thin disks. The slim disk hardly reduce to the thin disk in the presence of the secondary.% \vp{Is this true if say $\dot m=0.01$ or $\dot m=0.001$? Could it be that the two become the same at much lower accretion rates?}. \sw{yes, even mdot=0.001, slim disk does not reduce to thin disk, unless the q become very small, but reword "could not" as "hardly"}

At $r < a$, slim disks maintain a similar density profile even in the presence of the secondary, whereas thin disks exhibit a reduced density profile due to the influence of the secondary. This difference arises because the viscous parameter $\nu_0$ for thin disks are altered by the boundary conditions. If we instead fix the viscous parameter $\nu_0$  and allow the density to adjust to meet the boundary conditions, thin disks would develop a density profile similar to that of slim disks at $r<a$, under the presence of the secondary. However, this would result in  significantly larger surface density for $r > a$.

%\vp{Remove this discussion because the plot does not look very smooth?}The middle plot of Figure~\ref{fig:thin_vs_slim} shows the ratio \( Q^{\mathrm{tid}}/Q^{\mathrm{vis}} \) for slim and thin disks in the case of \( q=10^{-3} \). For all three cases, tidal torque heating dominates over viscous heating near \( r = a \), but decreases rapidly at both larger and smaller radii, consistent with the findings of \citet{Liu+10}. We note that the sharp increase in the ratio within the Hill sphere of the secondary for the slim disk is due to a significant drop in viscous torque, whereas both viscous heating and tidal torque heating decrease substantially within the secondary's Hill sphere. We also note that while \( Q^{\mathrm{tid}}/Q^{\mathrm{vis}} \) decreases rapidly beyond \( r > a \), \( Q^{\mathrm{rad}} \) remains substantially enhanced compared to the \( q = 0 \) case, even at large distances. This enhancement occurs because the tidal torque effectively amplifies the viscous torque (see the two left panels in the upper plot of Figure~\ref{fig:thin_vs_slim}).

The bottom plot of Figure~\ref{fig:thin_vs_slim} shows the spectra of slim and thin disks. For the cases with \( q = 0 \) (dotted color lines), both thin and slim disks have nearly identical spectra, indicating that the slim disk reduces to the thin disk at low accretion rates. For the cases with \( q = 10^{-3} \) (solid color lines), the thin disk spectra are much brighter than those of the slim disk, as well as brighter than the thin disk spectra for \( q = 0 \). Meanwhile, the slim disk spectrum is only slightly brighter than the slim disk spectrum for \( q = 0 \). As discussed above, this difference occurs because the thin disk has a different density profile than the slim disk, resulting in a much stronger integrated tidal torque.

\begin{figure}
    \centering    \includegraphics[width=0.45\textwidth]{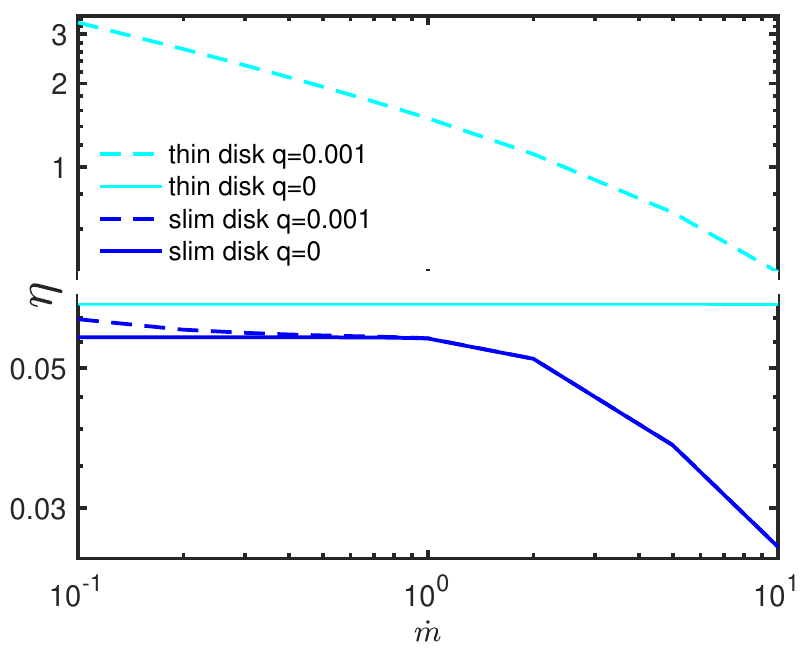}  %\includegraphics[width=0.45\textwidth]{thin_vs_slim_sp.pdf} 
    \caption{Radiation efficiency of thin disks versus slim disks. The disk parameters are set to $\alpha=0.1$, $M_p=10^8M_\odot$, and $a_d=100R_g$ for the slim disks. A break is inserted in the y-axis from 0.065 to 0.4 for better readability.}
    \label{fig:efficiency}
\end{figure}

Figure~\ref{fig:efficiency} shows the radiation efficiency $\eta$ of thin disks and slim disks, where $\eta$ is defined as $\eta = L_{\text{disk}}/(\dot{M} c^2)$. For the case of \( q = 0 \), \( \eta \sim 0.063 \) across all accretion rates, while \( \eta \) decreases from 0.055 to 0.025 as the accretion rate increases from sub-Eddington to significantly super-Eddington rates. The thin disk model assumes Keplerian orbits, with the disk terminating at the ISCO and cooling immediately via radiation. As a result, $\eta$ is insensitive to the accretion rate \citep{Shakura1973}. In contrast, the slim disk model does not require Keplerian orbits and allows advection to cool the disk, causing $\eta$ to decrease with increasing accretion rate \citep{Abramowicz1988, Sadowski2009}. From the plot, at low accretion rates ($\dot{m} \sim 0.1$), the thin disk is approximately 13\% more luminous than the slim disk, consistent with previous studies. For the case of \( q = 0.001 \), the thin disk exhibits \(\eta > 1\) at low accretion rates, indicating that most of the radiation originates from tidal heating. We note that such high $\eta$ values for a thin circumbinary disk can also be inferred from the results in \cite{Liu+10}. For example, using Table 1 of \cite{Liu+10}, we find that
$\eta$ increases from $\sim 0.1$ to $\sim3000$ for cases (b) to (d) of \cite{Liu+10}, where the accretion rate decreases from 0.15 to $4\times10^{-5}$. For slim disks, \(\eta\) increases only  slightly at low accretion rates, while at high accretion rates the efficiency of $q=0$ slim disks and $q>0$ slim disks are nearly identical. This suggests that tidal torque effect become negligible at high accretion rates, which is consistent with previous studies \citep{Narayan2000}.

\section{Slim disk results} \label{results}
In the section, we solve the binary slim disk equations numerically, and explore the impact of different parameters. We consider a binary with a primary of $M_{\bullet, p}=10^8 M_\odot$.

\subsection{General solutions}

\begin{figure}
    \centering    
    \includegraphics[width=0.45\textwidth]{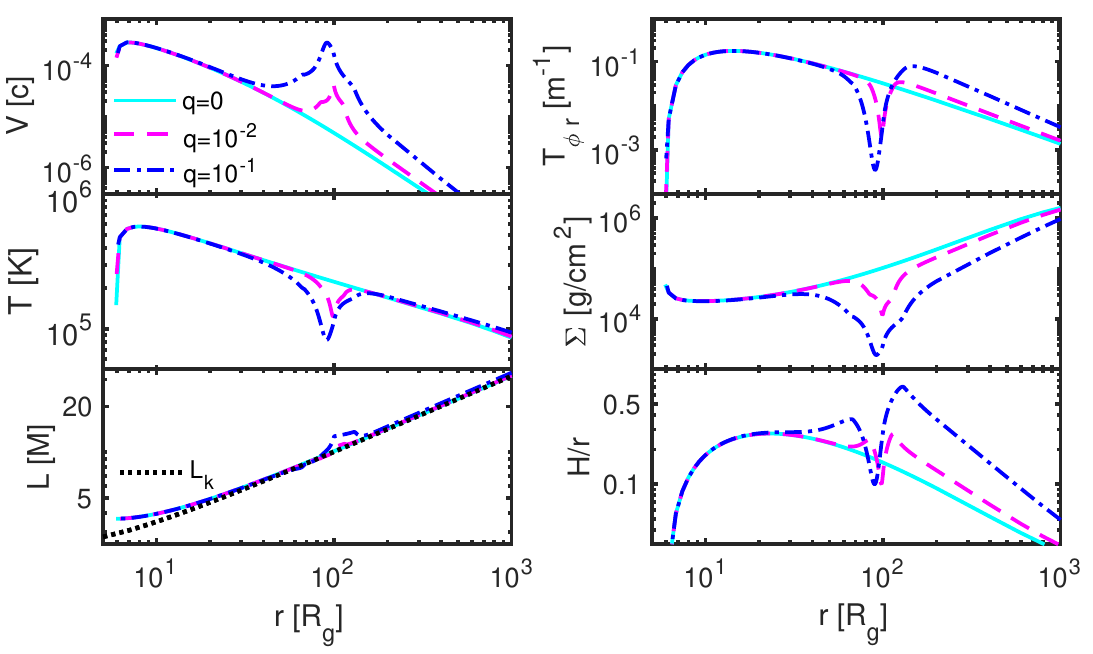}    
    \caption{
    Disk solutions of gas inflow velocity $V$, temperature $T$, specific angular momentum $L$, as well as viscous torque $T_{\phi r}$, surface density $\Sigma$ and scale high $H/r$, for cases of $q=0$ (cyan solid line), $q=0.01$ (magenta dashed line) and $q=0.1$ (blue dot dashed line). The other disk parameters are $\dot m=1$, $M_p=10^8M_\odot$, $a=100R_g$ and $\alpha=0.01$. }
    \label{fig:parameters}
\end{figure}

Figure \ref{fig:parameters} shows the disk solutions for the $q=0$, $q=0.01$, and $q=0.1$ cases. As $q$ increases, the tidal torque becomes stronger, producing a deeper dip in $T_{\phi r}$, $T$, $\Sigma$, and $H/r$ when $r\sim a$, while creating a stronger bump for the inflow velocity $V$. The strong tidal torque pushes gas away from the orbit, creating a gap in the disk with a reduced surface density even for slim disks. The decreased surface density is counteracted by a boost in the inflow velocity to keep the accretion rate constant. As seen in the upper right panel, the viscous torque becomes stronger at $r > a + R_H$, and then experiences a sharp drop at $r\sim a$. The drop in surface density is smaller than that in viscous torque. This is because the gas becomes cooler when both the viscous and the tidal torques decrease near $r\sim a$, resulting in a higher surface density due to vertical equilibrium. Near  $r\sim a$, gas 
specific angular momentum is super-Keplerian for $r > a$ and sub-Keplerian for $r < a$, since the secondary boosts the rotation of the outer disk and hinders the rotation of the inner disk, as shown in the lower left panel. 
The solutions for the inner part of the disk ($r<a$) are nearly the same for different values of $q$, as shown in Figure \ref{fig:parameters}. The integrated tidal torque does not significantly affect the inner part of the disk. As shown in Eq. \eqref{L_i}, $T_{\phi r}(r)$ is affected only by the tidal torque at $r_i < r$, where the tidal torque is small in the inner part of the disk. As a result, the secondary only slightly affects the thermal emission of the inner part of the disk.

\begin{figure}
    \centering    
    \includegraphics[width=0.45\textwidth]{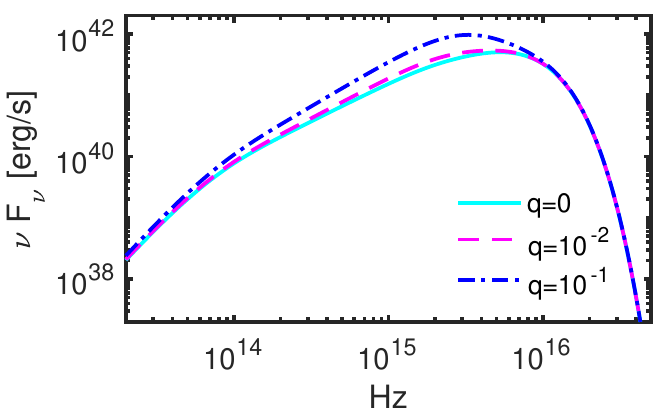}    
    \caption{The disk spectra on different q values. The other disk parameters are $\dot m=1$, $M_p=10^8M_\odot$, $a=100R_g$ and $\alpha=0.01$. }
    \label{fig:sp}
\end{figure}

Figure \ref{fig:sp} shows the spectra for cases of $q=0$, $q=10^{-2}$, and $q=0.1$. %We calculate the spectra by integrating the disk from $r=10,000R_g$ to ISCO. 
Since the emission from the inner part of the disk is only slightly affected by the tidal torque, all three cases exhibit nearly identical spectra at the blue end. However, since the tidal torque results in increased viscous torque at $r > a$, the outer disk becomes brighter as $q$ increases. As seen in the figure, the UV/optical emission becomes stronger with increasing $q$. Unlike thin disks \citep{Liu+10}, which modify both the inner and outer disk emission, the presence of the secondary only affects the outer disk emission ($r>a$), leaving the inner disk emission nearly unchanged. %

Disk emission provides a means to verify the model's consistency.
The bolometric luminosities of the three disks are $3.4 \times 10^{45}$, $3.8 \times 10^{45}$, and $6.2 \times 10^{45}$ ergs/s, indicating a $12\%$ and $82\%$ increase in brightness for the cases of $q=0.01$ and $q=0.1$, respectively, compared to the case of $q=0$. For simplicity, we assume that the orbital shrinkage of the secondary—which loses orbital energy—completely converts this energy into disk radiation. The additional flux contributed by the secondary (defined as the energy radiated due to its orbital evolution) can be approximated as \(2.9 \times 10^{46} qM_8 \, \text{erg/s} \)(approximately $2.3qL_{Edd}$). This estimate further assumes the secondary shrinks to the decoupling radius (\(a\sim200 R_g\)) from \(a\sim10,000 R_g\) within the shrinkage timescale($\sim5\times10^5$ years, see Sec.~\ref{ST} below), implying a radiation efficiency of about $0.25\%$ ($\Delta E= m c^2/2R_g\sim0.0025mc^2$). For $q = 0.01$ and $q = 0.1$, this would brighten the disk by approximately $8.5\%$ and $85\%$, respectively, relative to the bolometric luminosity of $3.4\times 10^{45}$ erg/s. These values align with the observed $12\%$ and $82\%$ increases in the binary slim disk solutions.

\subsection{The impact of $\alpha$, $\dot m$ and $a$}
\begin{figure}
    \centering    
    \includegraphics[width=0.45\textwidth]{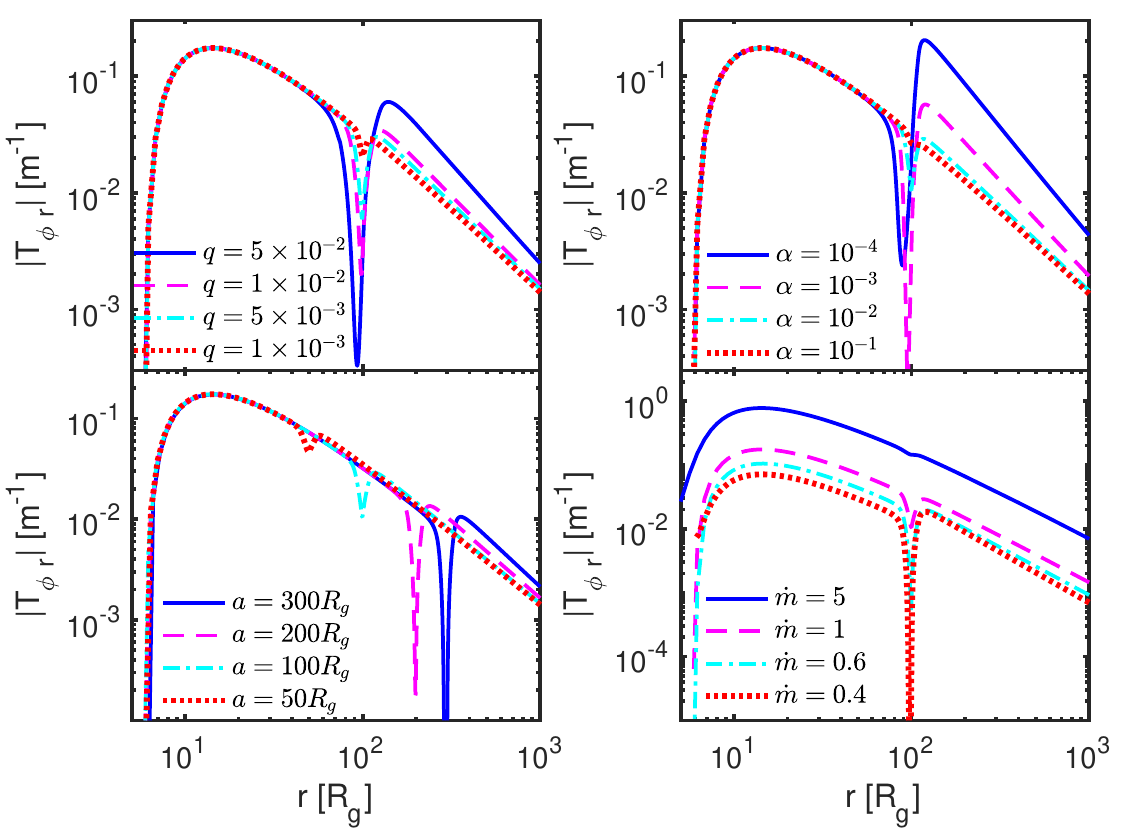}    
    \caption{
    The impact of varying $q$, $\alpha$, $a$, and $\dot{m}$ on the disk solution of $T_{\phi r}$. The chosen values for the fiducial disk parameters are $q=3\times10^{-3}$, $\alpha=0.01$, $M_{p}=10^8 M_\odot$, $a=100R_g$, and $\dot{m}=1$. In each panel, we illustrate the behavior of $T_{\phi r}$ as only a specific parameter is varied, while keeping all other disk parameters fixed at their fiducial values.}
    \label{fig:mal}
\end{figure}

Figure \ref{fig:mal} shows how the  viscous torque changes when varying parameters $q$, $\alpha$, $a$, and $\dot{m}$. A larger $q$ at fixed $a$, $\alpha$ and $\dot m$
results in a stronger integrated tidal torque, producing a deeper dip in $T_{\phi r}$ near $r\sim a$, and a stronger boost in $T_{\phi r}$ at $r > a$, as shown in the upper panel. A larger $a$ at fixed $q$, $\alpha$ and $\dot m$, has similar effect as shown in the lower left panel. This is because larger $a$ results in larger integrated tidal torque as demonstrated by our numerical results. Generally, a larger $a$ results in a smaller $\Lambda$, but amplifies the secondary's impact region and increases to total integrated torque. 
A smaller $\alpha$ at fixed $a$, $\dot m$ and $q$ results in a denser disk. Consequently, this yields a stronger integrated tidal torque, producing a deeper dip in $T_{\phi r}$ near $r=a$ and a stronger $T_{\phi r}$ at $r > a$, as shown in the upper right panels. For a smaller accretion rate at fixed $a$, $\alpha$ and $q$, the disk becomes thinner, leading to a more rapid decrease in the viscous torque compared to the integrated tidal torque \footnote{From vertical hydrostatic equilibrium, we have $P/\Sigma \propto (H/r)^2$. On the other hand, $H/r$ decreases when $\dot{m}$ decreases. Consequently, $P$ decreases more rapidly than $\Sigma$ when $\dot{m}$ decreases.}. Consequently, this results in a deeper dip in $T_{\phi r}$ near $a$ and a stronger $T_{\phi r}$ at $r > a$, as shown in the lower right panel. At super-Eddington accretion rates, the tidal torque can become negligible.

These results show that the tidal torque is highly sensitive to the viscous parameter, consistent with other studies \citep{Duffell2020,Dittmann2022,Wang2023,Turpin2024}. Studies on AGN indicate that $\alpha$ typically varies from $\sim0.1$ to $\sim0.001$ \citep{Brandenburg1995,Hawley1995,Starling2004}, while magnetized disks simulations show that $\alpha$ can depend on the geometry of the magnetic field and can vary at different radii \citep{Stone1996,Hirose2006,Hirose2009,Lan2019}. The strongest tidal torque occurs in systems with sub-Eddington accretion rate \footnote{This analysis focuses on an $\alpha$-disk. Advection-dominated disks \citep{Narayan1994}, characterized by accretion rates below a few percent of the Eddington limit, are not considered in this study.} and a large $q$. For some super-Eddington AGN, the tidal torque effect is indeed weak or sometimes negligible.

\subsection{The impacts of $U_{\rm s, r}$ and $U_{\rm s, z}$ }\label{Sec:impact_of_Usr_Usz}

\begin{figure}
    \centering    
    \includegraphics[width=0.43\textwidth]{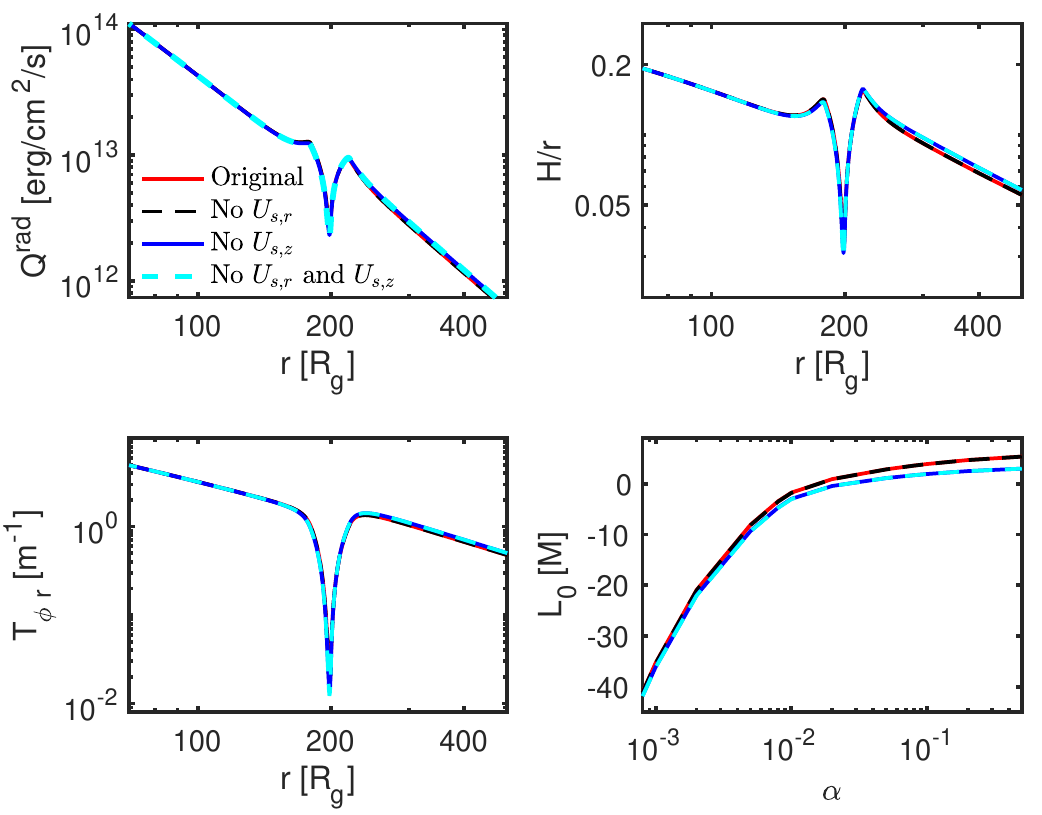}    
    \caption{ 
    The upper two and the lower left panels show the profiles of $V$, $H/r$ and $T_{\phi r}$, while the lower right panel shows the variation of $L_0$ as the viscous parameter $\alpha$ changes. The disk parameters are $M_{p}=10^8 M_\odot$, $\dot m=1$, $a=200R_g$, $q=0.003$ and $\alpha=0.01$.}
    \label{fig:usur}
\end{figure}

In this section, we explore the impact of the terms $U_{\rm s, r}$ and $U_{\rm s, z}$ by comparing the results with and without these terms included in the equations. Figure \ref{fig:usur} shows the effects of these two terms on the disk solutions.

As depicted in Figure \ref{fig:usur}, both $U_{\rm s, r}$ and  $U_{\rm s, z}$ have  limited impact on the disk radiative flux, and viscous torque (see the solid red and blue dashed lines). The solutions without $U_{\rm s, r}$ and  $U_{\rm s, z}$ are similar to the results including these terms (differing by $\sim3\%$). This occurs because the gravitational potential exerted on the gas by the primary is much stronger than that from the orbit-averaged potential of the secondary. For example, setting $r_a=1.05$ and $z_a=0$, the ratio $\bar U_{s}/ U_P=1.5q$.

Disregarding the term $U_{\rm s, r}$ and  $U_{\rm s, z}$ results in disk flux and viscous torque that remain nearly unchanged. By contrast, $U_{\rm s, z}$ has an impact on the disk thickness and integrated torque. The additional vertical gravity compresses the gas near the secondary, resulting in a thinner disk region (as seen in the upper right panel). The lower right panel demonstrates that neglecting $U_{\rm s, z}$ increases the integrated tidal torque. These results indicate that the vertical component simultaneously makes the disk thinner and less dense.  In general, the existence of $U_{\rm s, z}$ slightly mitigates the effect of the secondary, while both $U_{\rm s, r}$ and $U_{\rm s, z}$ have limited impact on the disk flux.

\subsection{Dependence of $L_0$ on different parameters}

\begin{figure}
    \centering    
    \includegraphics[width=0.43\textwidth]{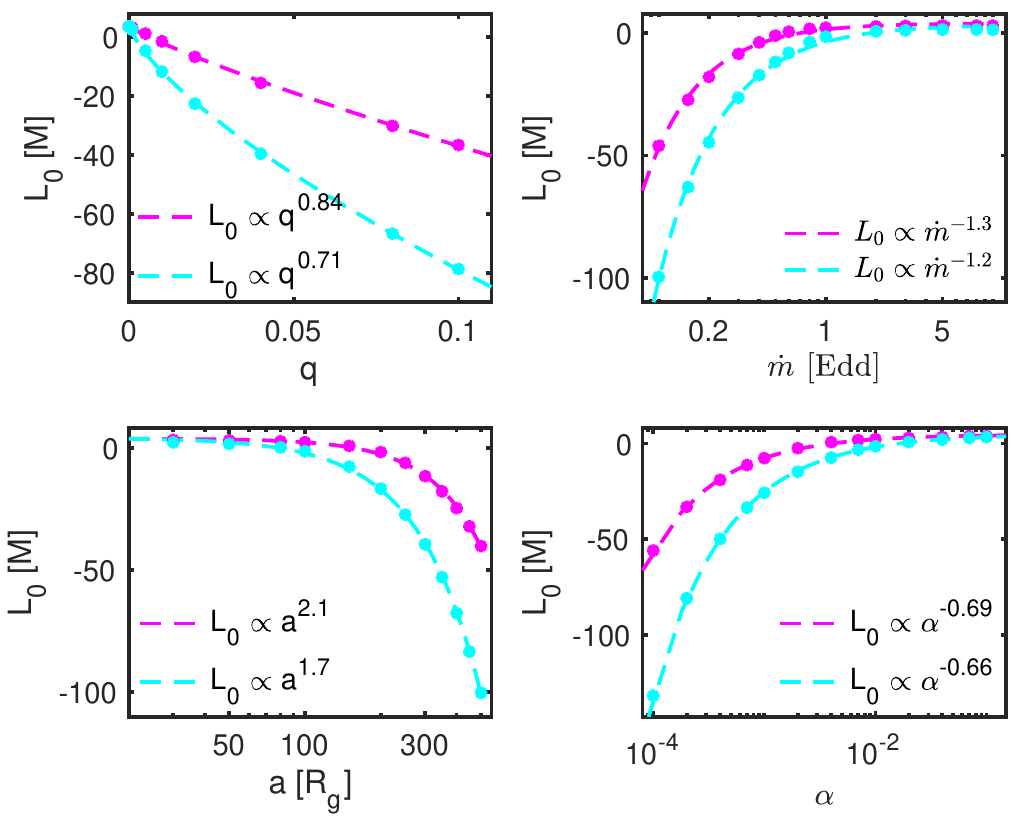}    
    \caption{Variation of $L_0$ with respect to different parameters. 
    In the upper-left panel, the magenta stars represent the cases with $\dot{m}=1$, while the cyan stars correspond to $\dot{m}=0.5$. In the other panels, the magenta stars indicate the case of $q=0.003$, whereas the cyan stars represent $q=0.01$. For all panels, the remaining disk parameters, not explicitly shown, are set to the following values: $\dot{m}=1$, $\alpha=0.01$, $a=100R_g$, and $q=0.003$.  The dashed lines are the best-fit power law model to the values. This figure show that $L_0$ is a power law function of $q$, $\dot m$, $a_d$ and $\alpha$, and the power law index is different for different parameters.}
    \label{fig:L_vs_p}
\end{figure}

$L_0$ is the sum of the angular momentum of the debris at the inner boundary ($L_{\rm i}$) and the total angular momentum transferred from the secondary to the debris, as defined in Eq. \eqref{L_0}. %
Since $L_{\rm i}$ remains nearly unchanged when varying $\alpha$ and $\dot m$ \citep{Abramowicz1988,Sadowski2009}, 
the behavior of $L_0$ reflects the behavior of the integrated tidal torque. The value of $L_0$ is a function of $q$, $a$, $\dot m$ and $\alpha$. The angular momentum of the debris transferred from the secondary are strongest for cases of sub-Eddington accretion rate and long separation.

Figure \ref{fig:L_vs_p} shows the dependence of $L_0$ on different parameters. The solid stars denote the values taken from the solutions, while the dashed lines are the best-fit power-law, $L_0=k_a X^{b}+L_i$. Here X denotes the specific parameters, $k_a$, $b$ and $L_i$ are free parameters that need to be fitted to the solution data. The coefficient of determination for the fit satisfies $R_d^2>0.995$ for the cases shown in the plot, indicating a good power-law fit to the data. $L_i$ changes slowly with different values of $\dot m$ and $\alpha$ as studied in the slim disk model \citep{Abramowicz1988,Sadowski2009}. The change of $L_0$ is mainly driven by the tidal torque and not a variation in $L_i$ (see Eq.~\eqref{L_0}), which can also be inferred from the good power-law fits.  The upper left panel shows the plot of $L_0$ VS $q$. The plot demonstrates that the smaller the accretion rate the steeper the decay of $L_0$ with increasing $q$. This is because the ratio of integrated tidal torque to viscous torque becomes larger in disks with lower accretion rates. The upper right and the lower two panels of Fig.~\ref{fig:L_vs_p}, show that the profile of $L_0$ vs the plotted parameter changes faster for cases with larger $q$. This can be attributed to stronger tidal torque associated with larger $q$. Assuming $\Sigma \propto r$ for a radiation-dominated disk \citep{Abramowicz1988}, which our solutions approximately satisfy, and that the tidal torque main contribution comes from $r=a$, using Eq.~\eqref{L_0} we obtain $L_0\propto q^{2/3}a^{2}$ for $r<a$, and $L_0\propto qa^{2}$ for $r>a$, close to the values shown in the upper and lower left panels.

\subsection{The shrinkage timescale}
\label{ST}

The rate of migration of the secondary by the tidal force and the gravitational wave torque can be expressed as $\dot a= \dot a_{\rm tid}+\dot a_{\rm gw}$. The $\dot a_{\rm tid}$ can be calculated as,
\begin{eqnarray}
\label{Tt}
&\dot a_{\rm tid}&=\frac{dL}{dt}/\frac{dL}{da}=\frac{2a^{1/2}\dot M(L_0-L_i)}{qM_p^{3/2}G^{1/2}}. 
%&&=2.0\times10^{-4} q^{-1} M_8 \dot m (L_0-L_i)a^{1/2} (m/s) \\
%&&=4.3\times10^{-8}q^{-1} M_8 \dot m (L_0-L_i)a^{1/2} (R_g/yr) \\
%&t&=\frac{a^{0.5-b}-2000^{0.5-b}}{(0.5-b)\times 6.1k_a}q10^8 yr
\end{eqnarray}
In the second equation, we have used $dL/dt=(L_0-L_i) \dot M$. $\dot a_{\rm gw}$ is given by \cite{Peters1964}, 
\begin{equation}
\dot a_{\rm gw}=-\frac{64 G^3M_p^3q}{5c^5a^3(1+q)^2}\frac{1+\frac{73}{24}e^2+\frac{37}{96}e^4}{(1-e^2)^{7/2}},%=-\frac{3.8\times 10^{9}q}{(a/R_g)^{3} (1+q)^2} (m/s)
\end{equation}
where $e$ is the orbital eccentricity. Using the relationship $L_0=K_aa^b+L_i$ as shown above, we are able to explore the evolution of the binary separation with time. Setting $\dot a_{\rm tid}=\dot a_{\rm gw}$, for a quasicircular orbit ($e=0$), we obtain the decoupling radius as,
\begin{equation}
%R_{\rm dec}=\left(-\frac{1.1\times 10^{13}q^2}{\dot m M_8 K_a(1+q)^2}\right)^{1/(b+3.5)}(R_g).
R_{\rm dec}=74^{5.5/(b+3.5)}\left(\frac{10^{-3}}{-K_a}\frac{(q/10^{-3})^2}{\dot m M_8 (1+q)^2}\right)^{1/(b+3.5)}(R_g).
\end{equation}
%\vp{What are the units of $R_{\rm dec}$? Can we plug in some values to get a sense of physical scale?}
In deriving this equation, we multiplied $(L_0 - L_i)$ by the factor $GM_p/c$ from the power-law fit to maintain consistent units. At $r< R_{\rm dec}$, gravitational radiation is the dominant process leading to orbital decay. As the outer parts of AGN disk can be gravitationally unstable beyond $\sim0.01-0.1$ pc \citep{Kolykhalov1979,Lodato2003,King2007}, we start from $a=0.05$pc ($\sim 10,000R_g$), and solve $\dot a= \dot a_{\rm tid}+\dot a_{\rm gw}$ from $a=10,000R_g$ to $a=50R_g$, for different parameters of circular orbit.

\begin{figure}
    \centering    
    \includegraphics[width=0.45\textwidth]{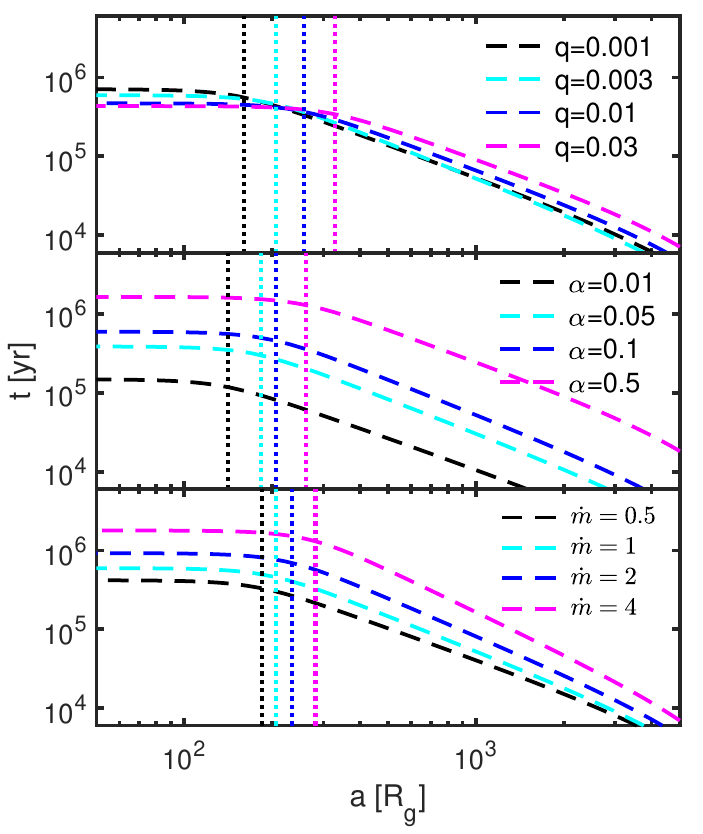}    
    \caption{%\vp{What is the primary mass here?} \sw{$M_p=10^8M_\odot$ added.} 
    Evolution of the binary separation with time. The disk parameters for the fiducial model are $M_p=10^8M_\odot$, $\dot m=1$, $\alpha=0.1$ and $q=3\times10^{-3}$. The vertical dotted lines show the critical radius where $\dot a_{\rm tid}=\dot a_{\rm gw}$ for the corresponding cases. This plot shows that the disk can help to shrink the separation effectively. }
    \label{fig:ta}
\end{figure}

Figure \ref{fig:ta} illustrates the evolution of the binary separation and the decoupling radii for different disk parameters. The decoupling radii are generally in the range of 100$R_g$ to 300$R_g$, depending on the strength of $\dot a_{\rm tid}$, consistent with the value of $\sim 100 R_g$ obtained by \cite{Armitage2002}. For the case of fixed $q$, $\dot a_{\rm tid}\propto L_{\rm 0}$. Consequently, weaker tidal torques result in a weaker $\dot a_{\rm tid}$, leading to a larger $R_{\rm dec}$. This can be seen in the middle and lower panels, where larger $\alpha$ and $\dot m$ produce larger $R_{\rm dec}$. However, for the case of varying $q$, $\dot a_{\rm tid}\propto q^{-1/3}$. As shown in the upper panel, a larger $q$ result in a larger critical radius, with even larger $q$ values leading to stronger tidal torques. This occurs because a black hole with greater mass must lose more angular momentum. 

The total shrinkage timescale is not highly sensitive to the mass ratio, as shown in the upper panel of Figure \ref{fig:ta}. This is because $\dot a_{\rm tid}\propto q^{-1/3}$, while $\dot a_{\rm gw}\propto q$, which reduces the dependence of $\dot a$ on $q$. In the middle panel, we observe that the shrinkage timescale increases as the viscous parameter $\alpha$ rises. This is because a higher $\alpha$ reduces the disk surface density, which in turn diminishes the tidal torque. The lower panel illustrates that the shrinkage timescale increases with the accretion rate, due to a substantial reduction in tidal torque \citep{Narayan2000}.  The shrinkage timescale is sensitive to  parameters, including $\alpha$ and $\dot m$. The time scale for the disk reducing the separation near the critical radius can change from approximately $0.1$ to $10$ million years for our fiducial binary with $M_p=10^8M_\odot$, consistent with some previous findings, for instance, \cite{Armitage2002MNRAS} demonstrated that a gaseous disk can shrink a binary system on a timescale of roughly $10^7$ years, and \cite{Robert2018} shown that type II migration would enable the secondary to shrink down the separation on a timescale shorter than the viscous timescale. However, it is inconsistent with some studies that suggest interaction with a finite gas disk is not an effective way to shrink the separation \citep{Lodato+09, Valli2024}.

%here we adopt $M_8\approx \sigma_{200}^4$ \citep{Tremaine2002}

%\begin{eqnarray}
%&&a = -1.69e-07, b = 2.94, c = 3.65 q=1e-3\\
%&&a = -7.2e-05, b = 2.1, c = 4.1 q=3e-3\\
%&&a = -3.86e-4, b = 1.92, c = 4.40 q=5e-3 \\
%&&a = -1.76e-3, b = 1.75, c = 4.78 q=1e-2 \\
%&&a = -7.86e-3, b = 1.62, c = 4.97 q=3e-2 \\
%&&a = -1.32e-2, b = 1.59, c = 4.80 q=5e-2 \\
%&&a = -2.01e-2, b = 1.57, c = 4.44 q=8e-2 
%\end{eqnarray}

%\subsection{The     disk thermal spectrum}

\section{Summary} \label{conclution}

We have developed a stationary circumbinary slim disk model through numerical simulations, solving the governing equations to characterize its structure. An analysis with the binary thin disk model \citep{Liu+10} was conducted to elucidate key differences in disk morphology and behavior. To quantify the influence of the secondary, we varied disk parameters—including $\alpha$, $\dot m$,  $q$ and $a$—and analyzed their effects on the equilibrium disk solutions and emitted spectra. We further evaluated the secondary’s perturbative effects by isolating contributions from its vertical and radial gravitational field components on the disk solutions. The integrated tidal torque exerted by the secondary on the disk was computed to estimate the orbital shrinkage timescale as a function of disk thermodynamics. These torque-driven calculations provide insights into the long-term orbital evolution of the binary system and its dependence on disk thermodynamics. Our findings are as follows:
\begin{enumerate}

\item 
In the presence of a secondary, a slim disk does not reduce to a thin disk even for low accretion rates. The tidal torque exerted on the slim disk is significantly smaller than that on the thin disk, due to the different disk density profiles.

\item
The tidal torque can boost the viscous torque at $r>a+R_H$, but it leaves nearly unchanged the inner parts of the disk, when compared to the solution of a single BH. This boosting effect makes the outer disk brighter, producing a signature characteristic of a binary disk.

\item 
The effect of tidal torque is highly sensitive to the disk parameters, including $q$, $\alpha$, $a$, and $\dot m$. Large $q$ increase the torque directly as indicated in Eq. \eqref{Lambda},% \vp{I don't see how increasing $a$ increases the torque. Maybe I messed up the math because I did it in this editor} \sw{move the $a$ below }, 
while larger $a$, smaller $\alpha$ and $\dot m$ result in a boost for the integrated tidal torque.

\item 
The vertical component of the gravity of the secondary slightly reduces the impact of the integrated tidal torque.  However, the vertical and the radial components of the secondary's gravitational field have a subdominant impact on the disk radiative flux.

\item 
The disk can effectively shrink the separation. The shrinkage timescale is highly sensitive to $\alpha$ and $\dot m$, but shows much weaker dependence on $q$.

\end{enumerate}

The slim disk model we presented in this work is suitable for systems with small mass ratio. Although the circumbinary slim disk is significantly less radiatively efficient, and hence dimmer at the same accretion rate, than its thin-disk counterpart, it confirms that tidal torque enhances emission in the outer disk regions at optical/UV wavelengths, producing an observational signature characteristic of binary systems. Some results from the circumbinary slim disk model are consistent with those of thin disks---for instance, the strength of tidal torque depends sensitively on several parameters, including \( q \), \( \alpha \), \( a \), and \( \dot{m} \) \citep{Duffell2020,Paschalidis:2021ntt,Bright:2022hnl,Dittmann2022,Wang2023,Turpin2024,Manikantan:2025afy}, while the influence of the secondary in the radial and vertical directions is negligible. We also find that the disk can effectively drive binary orbital shrinkage. However, these conclusions rely on assumptions of stationarity, circular orbits, and an infinite disk mass. For example, finite-mass eccentric disks may yield different outcomes \citep{Lodato+09,Valli2024}. A deeper understanding of binary disk dynamics will require further studies---particularly of essential, time-dependent, finite-mass models that ultimately include full general relativity and radiation feedback.

\begin{acknowledgements}
SW is supported by the NSFC (grant 12333004 and 12433005), the Strategic Priority Research Program of the Chinese Academy of Sciences (grant  XDB0550200), and the Strategic Pioneer Program on Space Science, Chinese Academy of Sciences (grant XDA15310300). SW thanks the UA Department of Astronomy and RU Department of Astrophysics/IMAPP for postdoctoral support during the early part of this work. This work was in part supported by NASA grant 80NSSC24K0771 and NSF grant PHY-2145421 to the University of Arizona, as well ACCESS award PHY190020.
\end{acknowledgements}

\begin{appendix}
\label{app:1}

\section{Impact of Different Vertical Integration Assumptions}

\begin{figure}[H]
   \centering    
    \includegraphics[width=0.45\textwidth]{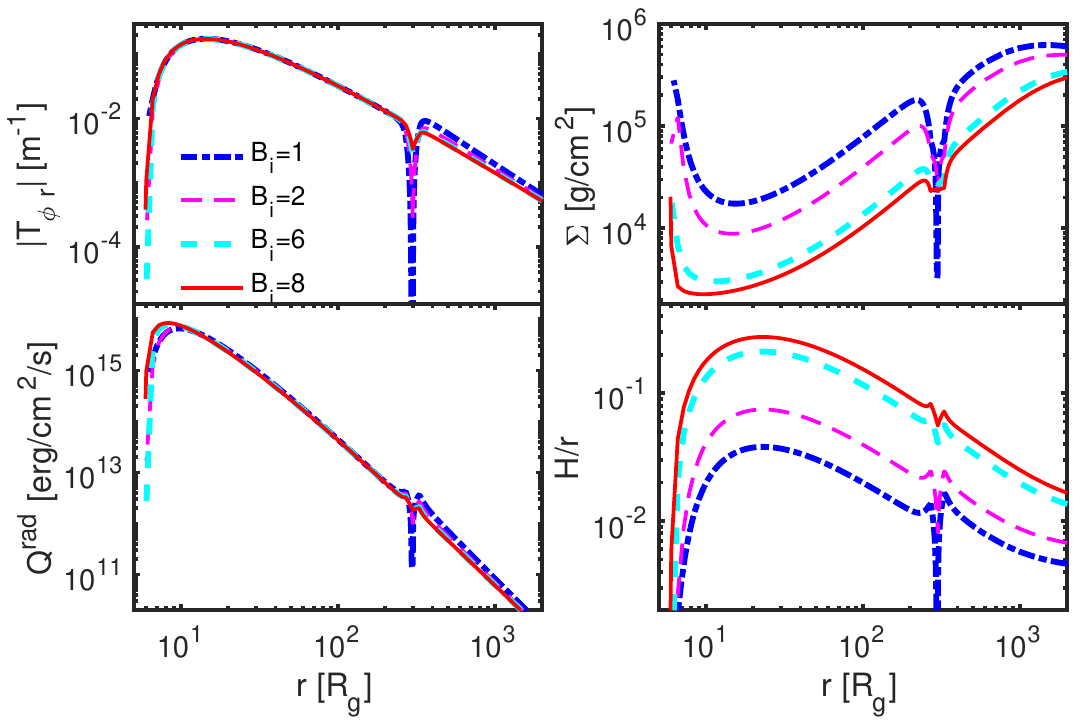}    
    \caption{The impact of varying $B_i$ on the disk solution. The chosen values for the disk parameters are $q=3\times10^{-3}$, $\alpha=0.1$, $M_{p}=10^8 M_\odot$, $a=300R_g$, and $\dot{m}=1$.}
    \label{fig:Bi}
\end{figure}

In this section, we explore the impact of different vertical integration assumptions. As shown in Eq.~(\ref{H2}), different assumptions result in different values of $B_i$. Figure~\ref{fig:Bi} illustrates the disk properties for varying $B_i$.

For the inner part of the disk ($r < a$), the viscous torque and disk radiation---unaffected by tidal torque---are insensitive to $B_i$. This indicates that the disk emission does not depend significantly on the vertical integration assumption, consistent with previous studies \citep{Abramowicz1988,sadowski2011}.

However, smaller $B_i$ values produce a thinner disk with lower density, as shown in the upper and lower right panels of Figure~\ref{fig:Bi}. The vertical integration assumption significantly affects the disk's thickness and density profile. Since the integrated tidal torque depends on density, smaller $B_i$ leads to a stronger integrated tidal torque. Consequently, $B_i$ affects the integrated tidal torque in the circumbinary disk, altering the emission from its outer region (see the upper and lower left panels). 

\end{appendix}

\bibliography{bib}{}

\begin{thebibliography}{}
\expandafter\ifx\csname natexlab\endcsname\relax\def\natexlab#1{#1}\fi
\providecommand{\url}[1]{\href{#1}{#1}}
\providecommand{\dodoi}[1]{doi:~\href{http://doi.org/#1}{\nolinkurl{#1}}}
\providecommand{\doeprint}[1]{\href{http://ascl.net/#1}{\nolinkurl{http://ascl.net/#1}}}
\providecommand{\doarXiv}[1]{\href{https://arxiv.org/abs/#1}{\nolinkurl{https://arxiv.org/abs/#1}}}

\bibitem[{{Abbott} {et~al.}(2016){Abbott}, {Abbott}, {Abbott}, {Abernathy},
  {Acernese}, {Ackley}, {Adams}, {Adams}, {Addesso}, {Adhikari}, {Adya},
  {Affeldt}, {Agathos}, {Agatsuma}, {Aggarwal}, {Aguiar}, {Aiello}, {Ain},
  {Ajith}, {Allen}, {Allocca}, {Altin}, {Anderson}, {Anderson}, {Arai},
  {Araya}, {Arceneaux}, {Areeda}, {Arnaud}, {Arun}, {Ascenzi}, {Ashton}, {Ast},
  {Aston}, {Astone}, {Aufmuth}, {Aulbert}, {Babak}, {Bacon}, {Bader}, {Baker},
  {Baldaccini}, {Ballardin}, {Ballmer}, {Barayoga}, {Barclay}, {Barish},
  {Barker}, {Barone}, {Barr}, {Barsotti}, {Barsuglia}, {Barta}, {Bartlett},
  {Bartos}, {Bassiri}, {Basti}, {Batch}, {Baune}, {Bavigadda}, {Bazzan},
  {Behnke}, {Bejger}, {Bell}, {Bell}, {Berger}, {Bergman}, {Bergmann}, {Berry},
  {Bersanetti}, {Bertolini}, {Betzwieser}, {Bhagwat}, {Bhandare}, {Bilenko},
  {Billingsley}, {Birch}, {Birney}, {Biscans}, {Bisht}, {Bitossi}, {Biwer},
  {Bizouard}, {Blackburn}, {Blair}, {Blair}, {Blair}, {Bloemen}, {Bock},
  {Bodiya}, {Boer}, {Bogaert}, {Bogan}, {Bohe}, {Bojtos}, {Bond}, {Bondu},
  {Bonnand}, {Boom}, {Bork}, {Boschi}, {Bose}, {Bouffanais}, {Bozzi},
  {Bradaschia}, {Brady}, {Braginsky}, {Branchesi}, {Brau}, {Briant}, {Brillet},
  {Brinkmann}, {Brisson}, {Brockill}, {Brooks}, {Brown}, {Brown}, {Brown},
  {Buchanan}, {Buikema}, {Bulik}, {Bulten}, {Buonanno}, {Buskulic}, {Buy},
  {Byer}, {Cadonati}, {Cagnoli}, {Cahillane}, {Calder{\'o}n Bustillo},
  {Callister}, {Calloni}, {Camp}, {Cannon}, {Cao}, {Capano}, {Capocasa},
  {Carbognani}, {Caride}, {Casanueva Diaz}, {Casentini}, {Caudill},
  {Cavagli{\`a}}, {Cavalier}, {Cavalieri}, {Cella}, {Cepeda}, {Cerboni
  Baiardi}, {Cerretani}, {Cesarini}, {Chakraborty}, {Chalermsongsak},
  {Chamberlin}, {Chan}, {Chao}, {Charlton}, {Chassande-Mottin}, {Chen}, {Chen},
  {Cheng}, {Chincarini}, {Chiummo}, {Cho}, {Cho}, {Chow}, {Christensen}, {Chu},
  {Chua}, {Chung}, {Ciani}, {Clara}, {Clark}, {Cleva}, {Coccia}, {Cohadon},
  {Colla}, {Collette}, {Cominsky}, {Constancio}, {Conte}, {Conti}, {Cook},
  {Corbitt}, {Cornish}, {Corsi}, {Cortese}, {Costa}, {Coughlin}, {Coughlin},
  {Coulon}, {Countryman}, {Couvares}, {Cowan}, {Coward}, {Cowart}, {Coyne},
  {Coyne}, {Craig}, {Creighton}, {Cripe}, {Crowder}, {Cumming}, {Cunningham},
  {Cuoco}, {Dal Canton}, {Danilishin}, {D'Antonio}, {Danzmann}, {Darman},
  {Dattilo}, {Dave}, {Daveloza}, {Davier}, {Davies}, {Daw}, {Day}, {DeBra},
  {Debreczeni}, {Degallaix}, {De Laurentis}, {Del{\'e}glise}, {Del Pozzo},
  {Denker}, {Dent}, {Dereli}, {Dergachev}, {DeRosa}, {De Rosa}, {DeSalvo},
  {Dhurandhar}, {D{\'\i}az}, {Di Fiore}, {Di Giovanni}, {Di Lieto}, {Di Pace},
  {Di Palma}, {Di Virgilio}, {Dojcinoski}, {Dolique}, {Donovan}, {Dooley},
  {Doravari}, {Douglas}, {Downes}, {Drago}, {Drever}, {Driggers}, {Du},
  {Ducrot}, {Dwyer}, {Edo}, {Edwards}, {Effler}, {Eggenstein}, {Ehrens},
  {Eichholz}, {Eikenberry}, {Engels}, {Essick}, {Etzel}, {Evans}, {Evans},
  {Everett}, {Factourovich}, {Fafone}, {Fair}, {Fairhurst}, {Fan}, {Fang},
  {Farinon}, {Farr}, {Farr}, {Favata}, {Fays}, {Fehrmann}, {Fejer}, {Ferrante},
  {Ferreira}, {Ferrini}, {Fidecaro}, {Fiori}, {Fiorucci}, {Fisher}, {Flaminio},
  {Fletcher}, {Fournier}, {Franco}, {Frasca}, {Frasconi}, {Frei}, {Freise},
  {Frey}, {Frey}, {Fricke}, {Fritschel}, {Frolov}, {Fulda}, {Fyffe}, {Gabbard},
  {Gair}, {Gammaitoni}, {Gaonkar}, {Garufi}, {Gatto}, {Gaur}, {Gehrels},
  {Gemme}, {Gendre}, {Genin}, {Gennai}, {George}, {Gergely}, {Germain},
  {Ghosh}, {Ghosh}, {Giaime}, {Giardina}, {Giazotto}, {Gill}, {Glaefke},
  {Goetz}, {Goetz}, {Gondan}, {Gonz{\'a}lez}, {Gonzalez Castro}, {Gopakumar},
  {Gordon}, {Gorodetsky}, {Gossan}, {Gosselin}, {Gouaty}, {Graef}, {Graff},
  {Granata}, {Grant}, {Gras}, {Gray}, {Greco}, {Green}, {Groot}, {Grote},
  {Grunewald}, {Guidi}, {Guo}, {Gupta}, {Gupta}, {Gushwa}, {Gustafson},
  {Gustafson}, {Hacker}, {Hall}, {Hall}, {Hammond}, {Haney}, {Hanke}, {Hanks},
  {Hanna}, {Hannam}, {Hanson}, {Hardwick}, {Haris}, {Harms}, {Harry}, {Harry},
  {Hart}, {Hartman}, {Haster}, {Haughian}, {Heidmann}, {Heintze}, {Heitmann},
  {Hello}, {Hemming}, {Hendry}, {Heng}, {Hennig}, {Heptonstall}, {Heurs},
  {Hild}, {Hoak}, {Hodge}, {Hofman}, {Hollitt}, {Holt}, {Holz}, {Hopkins},
  {Hosken}, {Hough}, {Houston}, {Howell}, {Hu}, {Huang}, {Huerta}, {Huet},
  {Hughey}, {Husa}, {Huttner}, {Huynh-Dinh}, {Idrisy}, {Indik}, {Ingram},
  {Inta}, {Isa}, {Isac}, {Isi}, {Islas}, {Isogai}, {Iyer}, {Izumi}, {Jacqmin},
  {Jang}, {Jani}, {Jaranowski}, {Jawahar}, {Jim{\'e}nez-Forteza}, {Johnson},
  {Jones}, {Jones}, {Jonker}, {Ju}, {Kalaghatgi}, {Kalogera}, {Kandhasamy},
  {Kang}, {Kanner}, {Karki}, {Kasprzack}, {Katsavounidis}, {Katzman}, {Kaufer},
  {Kaur}, {Kawabe}, {Kawazoe}, {K{\'e}f{\'e}lian}, {Kehl}, {Keitel}, {Kelley},
  {Kells}, {Kennedy}, {Key}, {Khalaidovski}, {Khalili}, {Khan}, {Khan}, {Khan},
  {Khazanov}, {Kijbunchoo}, {Kim}, {Kim}, {Kim}, {Kim}, {Kim}, {Kim}, {King},
  {King}, {Kinzel}, {Kissel}, {Kleybolte}, {Klimenko}, {Koehlenbeck},
  {Kokeyama}, {Koley}, {Kondrashov}, {Kontos}, {Korobko}, {Korth}, {Kowalska},
  {Kozak}, {Kringel}, {Kr{\'o}lak}, {Krueger}, {Kuehn}, {Kumar}, {Kuo},
  {Kutynia}, {Lackey}, {Landry}, {Lange}, {Lantz}, {Lasky}, {Lazzarini},
  {Lazzaro}, {Leaci}, {Leavey}, {Lebigot}, {Lee}, {Lee}, {Lee}, {Lee}, {Lenon},
  {Leonardi}, {Leong}, {Leroy}, {Letendre}, {Levin}, {Levine}, {Li}, {Libson},
  {Littenberg}, {Lockerbie}, {Logue}, {Lombardi}, {Lord}, {Lorenzini},
  {Loriette}, {Lormand}, {Losurdo}, {Lough}, {L{\"u}ck}, {Lundgren}, {Luo},
  {Lynch}, {Ma}, {MacDonald}, {Machenschalk}, {MacInnis}, {Macleod},
  {Maga{\~n}a-Sandoval}, {Magee}, {Mageswaran}, {Majorana}, {Maksimovic},
  {Malvezzi}, {Man}, {Mandel}, {Mandic}, {Mangano}, {Mansell}, {Manske},
  {Mantovani}, {Marchesoni}, {Marion}, {M{\'a}rka}, {M{\'a}rka}, {Markosyan},
  {Maros}, {Martelli}, {Martellini}, {Martin}, {Martin}, {Martynov}, {Marx},
  {Mason}, {Masserot}, {Massinger}, {Masso-Reid}, {Matichard}, {Matone},
  {Mavalvala}, {Mazumder}, {Mazzolo}, {McCarthy}, {McClelland}, {McCormick},
  {McGuire}, {McIntyre}, {McIver}, {McManus}, {McWilliams}, {Meacher},
  {Meadors}, {Meidam}, {Melatos}, {Mendell}, {Mendoza-Gandara}, {Mercer},
  {Merilh}, {Merzougui}, {Meshkov}, {Messenger}, {Messick}, {Meyers},
  {Mezzani}, {Miao}, {Michel}, {Middleton}, {Mikhailov}, {Milano}, {Miller},
  {Millhouse}, {Minenkov}, {Ming}, {Mirshekari}, {Mishra}, {Mitra},
  {Mitrofanov}, {Mitselmakher}, {Mittleman}, {Moggi}, {Mohan}, {Mohapatra},
  {Montani}, {Moore}, {Moore}, {Moraru}, {Moreno}, {Morriss}, {Mossavi},
  {Mours}, {Mow-Lowry}, {Mueller}, {Mueller}, {Muir}, {Mukherjee}, {Mukherjee},
  {Mukherjee}, {Mukund}, {Mullavey}, {Munch}, {Murphy}, {Murray}, {Mytidis},
  {Nardecchia}, {Naticchioni}, {Nayak}, {Necula}, {Nedkova}, {Nelemans},
  {Neri}, {Neunzert}, {Newton}, {Nguyen}, {Nielsen}, {Nissanke}, {Nitz},
  {Nocera}, {Nolting}, {Normandin}, {Nuttall}, {Oberling}, {Ochsner}, {O'Dell},
  {Oelker}, {Ogin}, {Oh}, {Oh}, {Ohme}, {Oliver}, {Oppermann}, {Oram},
  {O'Reilly}, {O'Shaughnessy}, {Ottaway}, {Ottens}, {Overmier}, {Owen}, {Pai},
  {Pai}, {Palamos}, {Palashov}, {Palomba}, {Pal-Singh}, {Pan}, {Pankow},
  {Pannarale}, {Pant}, {Paoletti}, {Paoli}, {Papa}, {Paris}, {Parker},
  {Pascucci}, {Pasqualetti}, {Passaquieti}, {Passuello}, {Patricelli},
  {Patrick}, {Pearlstone}, {Pedraza}, {Pedurand}, {Pekowsky}, {Pele}, {Penn},
  {Perreca}, {Phelps}, {Piccinni}, {Pichot}, {Piergiovanni}, {Pierro},
  {Pillant}, {Pinard}, {Pinto}, {Pitkin}, {Poggiani}, {Popolizio}, {Post},
  {Powell}, {Prasad}, {Predoi}, {Premachandra}, {Prestegard}, {Price},
  {Prijatelj}, {Principe}, {Privitera}, {Prodi}, {Prokhorov}, {Puncken},
  {Punturo}, {Puppo}, {P{\"u}rrer}, {Qi}, {Qin}, {Quetschke}, {Quintero},
  {Quitzow-James}, {Raab}, {Rabeling}, {Radkins}, {Raffai}, {Raja},
  {Rakhmanov}, {Rapagnani}, {Raymond}, {Razzano}, {Re}, {Read}, {Reed},
  {Regimbau}, {Rei}, {Reid}, {Reitze}, {Rew}, {Reyes}, {Ricci}, {Riles},
  {Robertson}, {Robie}, {Robinet}, {Rocchi}, {Rolland}, {Rollins}, {Roma},
  {Romano}, {Romanov}, {Romie}, {Rosi{\'n}ska}, {Rowan}, {R{\"u}diger},
  {Ruggi}, {Ryan}, {Sachdev}, {Sadecki}, {Sadeghian}, {Salconi}, {Saleem},
  {Salemi}, {Samajdar}, {Sammut}, {Sanchez}, {Sandberg}, {Sandeen}, {Sanders},
  {Sassolas}, {Sathyaprakash}, {Saulson}, {Sauter}, {Savage}, {Sawadsky},
  {Schale}, {Schilling}, {Schmidt}, {Schmidt}, {Schnabel}, {Schofield},
  {Sch{\"o}nbeck}, {Schreiber}, {Schuette}, {Schutz}, {Scott}, {Scott},
  {Sellers}, {Sengupta}, {Sentenac}, {Sequino}, {Sergeev}, {Serna},
  {Setyawati}, {Sevigny}, {Shaddock}, {Shah}, {Shahriar}, {Shaltev}, {Shao},
  {Shapiro}, {Shawhan}, {Sheperd}, {Shoemaker}, {Shoemaker}, {Siellez},
  {Siemens}, {Sigg}, {Silva}, {Simakov}, {Singer}, {Singer}, {Singh}, {Singh},
  {Singhal}, {Sintes}, {Slagmolen}, {Smith}, {Smith}, {Smith}, {Son}, {Sorazu},
  {Sorrentino}, {Souradeep}, {Srivastava}, {Staley}, {Steinke}, {Steinlechner},
  {Steinlechner}, {Steinmeyer}, {Stephens}, {Stone}, {Strain}, {Straniero},
  {Stratta}, {Strauss}, {Strigin}, {Sturani}, {Stuver}, {Summerscales}, {Sun},
  {Sutton}, {Swinkels}, {Szczepa{\'n}czyk}, {Tacca}, {Talukder}, {Tanner},
  {T{\'a}pai}, {Tarabrin}, {Taracchini}, {Taylor}, {Theeg},
  {Thirugnanasambandam}, {Thomas}, {Thomas}, {Thomas}, {Thorne}, {Thorne},
  {Thrane}, {Tiwari}, {Tiwari}, {Tokmakov}, {Tomlinson}, {Tonelli}, {Torres},
  {Torrie}, {T{\"o}yr{\"a}}, {Travasso}, {Traylor}, {Trifir{\`o}}, {Tringali},
  {Trozzo}, {Tse}, {Turconi}, {Tuyenbayev}, {Ugolini}, {Unnikrishnan}, {Urban},
  {Usman}, {Vahlbruch}, {Vajente}, {Valdes}, {van Bakel}, {van Beuzekom}, {van
  den Brand}, {Van Den Broeck}, {Vander-Hyde}, {van der Schaaf}, {van
  Heijningen}, {van Veggel}, {Vardaro}, {Vass}, {Vas{\'u}th}, {Vaulin},
  {Vecchio}, {Vedovato}, {Veitch}, {Veitch}, {Venkateswara}, {Verkindt},
  {Vetrano}, {Vicer{\'e}}, {Vinciguerra}, {Vine}, {Vinet}, {Vitale}, {Vo},
  {Vocca}, {Vorvick}, {Voss}, {Vousden}, {Vyatchanin}, {Wade}, {Wade}, {Wade},
  {Walker}, {Wallace}, {Walsh}, {Wang}, {Wang}, {Wang}, {Wang}, {Wang}, {Ward},
  {Warner}, {Was}, {Weaver}, {Wei}, {Weinert}, {Weinstein}, {Weiss}, {Welborn},
  {Wen}, {We{\ss}els}, {Westphal}, {Wette}, {Whelan}, {Whitcomb}, {White},
  {Whiting}, {Williams}, {Williamson}, {Willis}, {Willke}, {Wimmer}, {Winkler},
  {Wipf}, {Wittel}, {Woan}, {Worden}, {Wright}, {Wu}, {Yablon}, {Yam},
  {Yamamoto}, {Yancey}, {Yap}, {Yu}, {Yvert}, {Zadro{\.Z}ny}, {Zangrando},
  {Zanolin}, {Zendri}, {Zevin}, {Zhang}, {Zhang}, {Zhang}, {Zhang}, {Zhao},
  {Zhou}, {Zhou}, {Zhu}, {Zucker}, {Zuraw}, {Zweizig}, {LIGO Scientific
  Collaboration}, \& {Virgo Collaboration}}]{Abbott2016}
{Abbott}, B.~P., {Abbott}, R., {Abbott}, T.~D., {et~al.} 2016, \prl, 116,
  131103, \dodoi{10.1103/PhysRevLett.116.131103}

\bibitem[{{Abbott} {et~al.}(2023){Abbott}, {Abbott}, {Acernese}, {Ackley},
  {Adams}, {Adhikari}, {Adhikari}, {Adya}, {Affeldt}, {Agarwal}, {Agathos},
  {Agatsuma}, {Aggarwal}, {Aguiar}, {Aiello}, {Ain}, {Ajith}, {Akcay},
  {Akutsu}, {Albanesi}, {Allocca}, {Altin}, {Amato}, {Anand}, {Anand},
  {Ananyeva}, {Anderson}, {Anderson}, {Ando}, {Andrade}, {Andres},
  {Andri{\'c}}, {Angelova}, {Ansoldi}, {Antelis}, {Antier}, {Appert}, {Arai},
  {Arai}, {Arai}, {Araki}, {Araya}, {Araya}, {Areeda}, {Ar{\`e}ne}, {Aritomi},
  {Arnaud}, {Arogeti}, {Aronson}, {Arun}, {Asada}, {Asali}, {Ashton}, {Aso},
  {Assiduo}, {Aston}, {Astone}, {Aubin}, {Austin}, {Babak}, {Badaracco},
  {Bader}, {Badger}, {Bae}, {Bae}, {Baer}, {Bagnasco}, {Bai}, {Baiotti},
  {Baird}, {Bajpai}, {Ball}, {Ballardin}, {Ballmer}, {Balsamo}, {Baltus},
  {Banagiri}, {Bankar}, {Barayoga}, {Barbieri}, {Barish}, {Barker}, {Barneo},
  {Barone}, {Barr}, {Barsotti}, {Barsuglia}, {Barta}, {Bartlett}, {Barton},
  {Bartos}, {Bassiri}, {Basti}, {Bawaj}, {Bayley}, {Baylor}, {Bazzan},
  {B{\'e}csy}, {Bedakihale}, {Bejger}, {Belahcene}, {Benedetto}, {Beniwal},
  {Bennett}, {Bentley}, {Benyaala}, {Bergamin}, {Berger}, {Bernuzzi}, {Berry},
  {Bersanetti}, {Bertolini}, {Betzwieser}, {Beveridge}, {Bhandare}, {Bhardwaj},
  {Bhattacharjee}, {Bhaumik}, {Bilenko}, {Billingsley}, {Bini}, {Birney},
  {Birnholtz}, {Biscans}, {Bischi}, {Biscoveanu}, {Bisht}, {Biswas}, {Bitossi},
  {Bizouard}, {Blackburn}, {Blair}, {Blair}, {Blair}, {Bobba}, {Bode}, {Boer},
  {Bogaert}, {Boldrini}, {Bonavena}, {Bondu}, {Bonilla}, {Bonnand}, {Booker},
  {Boom}, {Bork}, {Boschi}, {Bose}, {Bose}, {Bossilkov}, {Boudart},
  {Bouffanais}, {Bozzi}, {Bradaschia}, {Brady}, {Bramley}, {Branch},
  {Branchesi}, {Brandt}, {Brau}, {Breschi}, {Briant}, {Briggs}, {Brillet},
  {Brinkmann}, {Brockill}, {Brooks}, {Brooks}, {Brown}, {Brunett}, {Bruno},
  {Bruntz}, {Bryant}, {Bulik}, {Bulten}, {Buonanno}, {Buscicchio}, {Buskulic},
  {Buy}, {Byer}, {Davies}, {Cadonati}, {Cagnoli}, {Cahillane}, {Bustillo},
  {Callaghan}, {Callister}, {Calloni}, {Cameron}, {Camp}, {Canepa},
  {Canevarolo}, {Cannavacciuolo}, {Cannon}, {Cao}, {Cao}, {Capocasa}, {Capote},
  {Carapella}, {Carbognani}, {Carlin}, {Carney}, {Carpinelli}, {Carrillo},
  {Carullo}, {Carver}, {Diaz}, {Casentini}, {Castaldi}, {Caudill},
  {Cavagli{\`a}}, {Cavalier}, {Cavalieri}, {Ceasar}, {Cella},
  {Cerd{\'a}-Dur{\'a}n}, {Cesarini}, {Chaibi}, {Chakravarti}, {Subrahmanya},
  {Champion}, {Chan}, {Chan}, {Chan}, {Chan}, {Chan}, {Chandra}, {Chanial},
  {Chao}, {Chapman-Bird}, {Charlton}, {Chase}, {Chassande-Mottin},
  {Chatterjee}, {Chatterjee}, {Chatterjee}, {Chaturvedi}, {Chaty},
  {Chatziioannou}, {Chen}, {Chen}, {Chen}, {Chen}, {Chen}, {Chen}, {Chen},
  {Chen}, {Cheng}, {Cheong}, {Cheung}, {Chia}, {Chiadini}, {Chiang},
  {Chiarini}, {Chierici}, {Chincarini}, {Chiofalo}, {Chiummo}, {Cho}, {Cho},
  {Choudhary}, {Choudhary}, {Christensen}, {Chu}, {Chu}, {Chu}, {Chua},
  {Chung}, {Ciani}, {Ciecielag}, {Cie{\'s}lar}, {Cifaldi}, {Ciobanu}, {Ciolfi},
  {Cipriano}, {Cirone}, {Clara}, {Clark}, {Clark}, {Clarke}, {Clearwater},
  {Clesse}, {Cleva}, {Coccia}, {Codazzo}, {Cohadon}, {Cohen}, {Cohen},
  {Colleoni}, {Collette}, {Colombo}, {Colpi}, {Compton}, {Constancio}, {Conti},
  {Cooper}, {Corban}, {Corbitt}, {Cordero-Carri{\'o}n}, {Corezzi}, {Corley},
  {Cornish}, {Corre}, {Corsi}, {Cortese}, {Costa}, {Cotesta}, {Coughlin},
  {Coulon}, {Countryman}, {Cousins}, {Couvares}, {Coward}, {Cowart}, {Coyne},
  {Coyne}, {Creighton}, {Creighton}, {Criswell}, {Croquette}, {Crowder},
  {Cudell}, {Cullen}, {Cumming}, {Cummings}, {Cunningham}, {Cuoco},
  {Cury{\l}o}, {Dabadie}, {Canton}, {Dall'Osso}, {D{\'a}lya}, {Dana},
  {Daneshgaranbajastani}, {D'Angelo}, {Danila}, {Danilishin}, {D'Antonio},
  {Danzmann}, {Darsow-Fromm}, {Dasgupta}, {Datrier}, {Dattilo}, {Dave},
  {Davier}, {Davis}, {Davis}, {Daw}, {de Alarc{\'o}n}, {Dean}, {Debra},
  {Deenadayalan}, {Degallaix}, {de Laurentis}, {Del{\'e}glise}, {Del Favero},
  {de Lillo}, {de Lillo}, {Del Pozzo}, {Demarchi}, {de Matteis}, {D'Emilio},
  {Demos}, {Dent}, {Depasse}, {de Pietri}, {De Rosa}, {de Rossi}, {Desalvo},
  {de Simone}, {Dhurandhar}, {D{\'\i}az}, {Diaz-Ortiz}, {Didio}, {Dietrich},
  {di Fiore}, {di Fronzo}, {di Giorgio}, {di Giovanni}, {di Giovanni}, {di
  Girolamo}, {di Lieto}, {Ding}, {di Pace}, {di Palma}, {di Renzo},
  {Divakarla}, {Dmitriev}, {Doctor}, {D'Onofrio}, {Donovan}, {Dooley},
  {Doravari}, {Dorrington}, {Drago}, {Driggers}, {Drori}, {Ducoin}, {Dupej},
  {Durante}, {D'Urso}, {Duverne}, {Dwyer}, {Eassa}, {Easter}, {Ebersold},
  {Eckhardt}, {Eddolls}, {Edelman}, {Edo}, {Edy}, {Effler}, {Eguchi},
  {Eichholz}, {Eikenberry}, {Eisenmann}, {Eisenstein}, {Ejlli}, {Engelby},
  {Enomoto}, {Errico}, {Essick}, {Estell{\'e}s}, {Estevez}, {Etienne}, {Etzel},
  {Evans}, {Evans}, {Ewing}, {Fafone}, {Fair}, {Fairhurst}, {Farah}, {Farinon},
  {Farr}, {Farr}, {Farrow}, {Fauchon-Jones}, {Favaro}, {Favata}, {Fays},
  {Fazio}, {Feicht}, {Fejer}, {Fenyvesi}, {Ferguson}, {Fernandez-Galiana},
  {Ferrante}, {Ferreira}, {Fidecaro}, {Figura}, {Fiori}, {Fishbach}, {Fisher},
  {Fittipaldi}, {Fiumara}, {Flaminio}, {Floden}, {Fong}, {Font}, {Fornal},
  {Forsyth}, {Franke}, {Frasca}, {Frasconi}, {Frederick}, {Freed}, {Frei},
  {Freise}, {Frey}, {Fritschel}, {Frolov}, {Fronz{\'e}}, {Fujii}, {Fujikawa},
  {Fukunaga}, {Fukushima}, {Fulda}, {Fyffe}, {Gabbard}, {Gabella}, {Gadre},
  {Gair}, {Gais}, {Galaudage}, {Gamba}, {Ganapathy}, {Ganguly}, {Gao},
  {Gaonkar}, {Garaventa}, {Garc{\'\i}a}, {Garc{\'\i}a-N{\'u}{\~n}ez},
  {Garc{\'\i}a-Quir{\'o}s}, {Garufi}, {Gateley}, {Gaudio}, {Gayathri}, {Ge},
  {Gemme}, {Gennai}, {George}, {George}, {Gerberding}, {Gergely}, {Gewecke},
  {Ghonge}, {Ghosh}, {Ghosh}, {Ghosh}, {Ghosh}, {Giacomazzo}, {Giacoppo},
  {Giaime}, {Giardina}, {Gibson}, {Gier}, {Giesler}, {Giri}, {Gissi},
  {Glanzer}, {Gleckl}, {Godwin}, {Goetz}, {Goetz}, {Gohlke}, {Golomb},
  {Goncharov}, {Gonz{\'a}lez}, {Gopakumar}, {Gosselin}, {Gouaty}, {Gould},
  {Grace}, {Grado}, {Granata}, {Granata}, {Grant}, {Gras}, {Grassia}, {Gray},
  {Gray}, {Greco}, {Green}, {Green}, {Gretarsson}, {Gretarsson}, {Griffith},
  {Griffiths}, {Griggs}, {Grignani}, {Grimaldi}, {Grimm}, {Grote}, {Grunewald},
  {Gruning}, {Guerra}, {Guidi}, {Guimaraes}, {Guix{\'e}}, {Gulati}, {Guo},
  {Guo}, {Gupta}, {Gupta}, {Gupta}, {Gustafson}, {Gustafson}, {Guzman}, {Ha},
  {Haegel}, {Hagiwara}, {Haino}, {Halim}, {Hall}, {Hamilton}, {Hammond}, {Han},
  {Haney}, {Hanks}, {Hanna}, \& {Hannam}}]{Abbott2023}
{Abbott}, R., {Abbott}, T.~D., {Acernese}, F., {et~al.} 2023, Physical Review
  X, 13, 041039, \dodoi{10.1103/PhysRevX.13.041039}

\bibitem[{Abramowicz {et~al.}(1996)Abramowicz, Chen, Granath, \&
  Lasota}]{Abramowicz1996}
Abramowicz, M.~A., Chen, X.-M., Granath, M., \& Lasota, J.-P. 1996, The
  Astrophysical Journal, 471, 762, \dodoi{10.1086/178004}

\bibitem[{Abramowicz {et~al.}(1988)Abramowicz, Czerny, Lasota, \&
  Szuszkiewicz}]{Abramowicz1988}
Abramowicz, M.~A., Czerny, B., Lasota, J.~P., \& Szuszkiewicz, E. 1988,
  Astrophys. J.; (United States), 332, 646, \dodoi{10.1086/166683}

\bibitem[{{Amaro-Seoane} {et~al.}(2023){Amaro-Seoane}, {Andrews}, {Arca Sedda},
  {Askar}, {Baghi}, {Balasov}, {Bartos}, {Bavera}, {Bellovary}, {Berry},
  {Berti}, {Bianchi}, {Blecha}, {Blondin}, {Bogdanovi{\'c}}, {Boissier},
  {Bonetti}, {Bonoli}, {Bortolas}, {Breivik}, {Capelo}, {Caramete},
  {Cattorini}, {Charisi}, {Chaty}, {Chen}, {Chru{\'s}li{\'n}ska}, {Chua},
  {Church}, {Colpi}, {D'Orazio}, {Danielski}, {Davies}, {Dayal}, {De Rosa},
  {Derdzinski}, {Destounis}, {Dotti}, {Du{\c{t}}an}, {Dvorkin}, {Fabj},
  {Foglizzo}, {Ford}, {Fouvry}, {Franchini}, {Fragos}, {Fryer}, {Gaspari},
  {Gerosa}, {Graziani}, {Groot}, {Habouzit}, {Haggard}, {Haiman}, {Han},
  {Istrate}, {Johansson}, {Khan}, {Kimpson}, {Kokkotas}, {Kong}, {Korol},
  {Kremer}, {Kupfer}, {Lamberts}, {Larson}, {Lau}, {Liu}, {Lloyd-Ronning},
  {Lodato}, {Lupi}, {Ma}, {Maccarone}, {Mandel}, {Mangiagli}, {Mapelli},
  {Mathis}, {Mayer}, {McGee}, {McKernan}, {Miller}, {Mota}, {Mumpower},
  {Nasim}, {Nelemans}, {Noble}, {Pacucci}, {Panessa}, {Paschalidis}, {Pfister},
  {Porquet}, {Quenby}, {Ricarte}, {R{\"o}pke}, {Regan}, {Rosswog}, {Ruiter},
  {Ruiz}, {Runnoe}, {Schneider}, {Schnittman}, {Secunda}, {Sesana}, {Seto},
  {Shao}, {Shapiro}, {Sopuerta}, {Stone}, {Suvorov}, {Tamanini}, {Tamfal},
  {Tauris}, {Temmink}, {Tomsick}, {Toonen}, {Torres-Orjuela}, {Toscani},
  {Tsokaros}, {Unal}, {V{\'a}zquez-Aceves}, {Valiante}, {van Putten}, {van
  Roestel}, {Vignali}, {Volonteri}, {Wu}, {Younsi}, {Yu}, {Zane}, {Zwick},
  {Antonini}, {Baibhav}, {Barausse}, {Bonilla Rivera}, {Branchesi},
  {Branduardi-Raymont}, {Burdge}, {Chakraborty}, {Cuadra}, {Dage}, {Davis}, {de
  Mink}, {Decarli}, {Doneva}, {Escoffier}, {Gandhi}, {Haardt}, {Lousto},
  {Nissanke}, {Nordhaus}, {O'Shaughnessy}, {Portegies Zwart}, {Pound},
  {Schussler}, {Sergijenko}, {Spallicci}, {Vernieri}, \&
  {Vigna-G{\'o}mez}}]{Amaro-Seoane2023}
{Amaro-Seoane}, P., {Andrews}, J., {Arca Sedda}, M., {et~al.} 2023, Living
  Reviews in Relativity, 26, 2, \dodoi{10.1007/s41114-022-00041-y}

\bibitem[{{Armitage} \& {Bonnell}(2002)}]{Armitage2002MNRAS}
{Armitage}, P.~J., \& {Bonnell}, I.~A. 2002, \mnras, 330, L11,
  \dodoi{10.1046/j.1365-8711.2002.05213.x}

\bibitem[{{Armitage} \& {Natarajan}(2002)}]{Armitage2002}
{Armitage}, P.~J., \& {Natarajan}, P. 2002, \apjl, 567, L9,
  \dodoi{10.1086/339770}

\bibitem[{Avara {et~al.}(2024)Avara, Krolik, Campanelli, Noble, Bowen, \&
  Ryu}]{Avara:2023ztw}
Avara, M.~J., Krolik, J.~H., Campanelli, M., {et~al.} 2024, Astrophys. J., 974,
  242, \dodoi{10.3847/1538-4357/ad5bda}

\bibitem[{{Baker} {et~al.}(2019){Baker}, {Haiman}, {Rossi}, {Berger}, {Brandt},
  {Breedt}, {Breivik}, {Charisi}, {Derdzinski}, {D'Orazio}, {Ford}, {Greene},
  {Hill}, {Holley-Bockelmann}, {Key}, {Kocsis}, {Kupfer}, {Madau}, {Marsh},
  {McKernan}, {McWilliams}, {Natarajan}, {Nissanke}, {Noble}, {Phinney},
  {Ramsay}, {Schnittman}, {Sesana}, {Shoemaker}, {Stone}, {Toonen},
  {Trakhtenbrot}, {Vikhlinin}, \& {Volonteri}}]{lisa_multimessenger_whitepaper}
{Baker}, J., {Haiman}, Z., {Rossi}, E.~M., {et~al.} 2019, \baas, 51, 123,
  \dodoi{10.48550/arXiv.1903.04417}

\bibitem[{{Begelman} {et~al.}(1980){Begelman}, {Blandford}, \&
  {Rees}}]{Begelman1980}
{Begelman}, M.~C., {Blandford}, R.~D., \& {Rees}, M.~J. 1980, \nat, 287, 307,
  \dodoi{10.1038/287307a0}

\bibitem[{Bogdanovic {et~al.}(2022)Bogdanovic, Miller, \&
  Blecha}]{bogdanovic_bhb_review}
Bogdanovic, T., Miller, M.~C., \& Blecha, L. 2022, Living Reviews in
  Relativity, 25, 3, \dodoi{10.1007/s41114-022-00037-8}

\bibitem[{{Brandenburg} {et~al.}(1995){Brandenburg}, {Nordlund}, {Stein}, \&
  {Torkelsson}}]{Brandenburg1995}
{Brandenburg}, A., {Nordlund}, A., {Stein}, R.~F., \& {Torkelsson}, U. 1995,
  \apj, 446, 741, \dodoi{10.1086/175831}

\bibitem[{Bright \& Paschalidis(2023)}]{Bright:2022hnl}
Bright, J.~C., \& Paschalidis, V. 2023, Mon. Not. Roy. Astron. Soc., 520, 392,
  \dodoi{10.1093/mnras/stad091}

\bibitem[{{Cattorini} \& {Giacomazzo}(2024)}]{Cattorini2024}
{Cattorini}, F., \& {Giacomazzo}, B. 2024, Astroparticle Physics, 154, 102892,
  \dodoi{10.1016/j.astropartphys.2023.102892}

\bibitem[{Cattorini \& Giacomazzo(2024)}]{Cattorini:2023akr}
Cattorini, F., \& Giacomazzo, B. 2024, Astropart. Phys., 154, 102892,
  \dodoi{10.1016/j.astropartphys.2023.102892}

\bibitem[{{Chang}(2008)}]{Chang2008}
{Chang}, P. 2008, \apj, 684, 236, \dodoi{10.1086/590326}

\bibitem[{Clyburn \& Zrake(2024)}]{Clyburn:2024xyq}
Clyburn, M., \& Zrake, J. 2024.
\newblock \doarXiv{2405.10281}

\bibitem[{{Colpi} {et~al.}(2019){Colpi}, {Holley-Bockelmann}, {Bogdanovic},
  {Natarajan}, {Bellovary}, {Sesana}, {Tremmel}, {Schnittman}, {Comerford},
  {Barausse}, {Berti}, {Volonteri}, {Khan}, {McWilliams}, {Burke-Spolaor},
  {Hazboun}, {Conklin}, {Mueller}, \& {Larson}}]{lisa_gw_whitepaper}
{Colpi}, M., {Holley-Bockelmann}, K., {Bogdanovic}, T., {et~al.} 2019, arXiv
  e-prints, arXiv:1903.06867, \dodoi{10.48550/arXiv.1903.06867}

\bibitem[{{Corrales} {et~al.}(2010){Corrales}, {Haiman}, \&
  {MacFadyen}}]{Corrales2010}
{Corrales}, L.~R., {Haiman}, Z., \& {MacFadyen}, A. 2010, \mnras, 404, 947,
  \dodoi{10.1111/j.1365-2966.2010.16324.x}

\bibitem[{{Cox} \& {Giuli}(1968)}]{Cox1968}
{Cox}, J.~P., \& {Giuli}, R.~T. 1968, {Principles of stellar structure}

\bibitem[{DeLaurentiis {et~al.}(2024)DeLaurentiis, Haiman,
  Westernacher-Schneider, Krauth, Davelaar, Zrake, \&
  MacFadyen}]{delaurentiis_relativistic_2024}
DeLaurentiis, S., Haiman, Z., Westernacher-Schneider, J.~R., {et~al.} 2024,
  Relativistic {Binary} {Precession}: {Impact} on {Eccentric} {Binary}
  {Accretion} and {Multi}-{Messenger} {Astronomy},  arXiv.
\newblock \url{http://arxiv.org/abs/2405.07897}

\bibitem[{{Dittmann} \& {Ryan}(2022)}]{Dittmann2022}
{Dittmann}, A.~J., \& {Ryan}, G. 2022, \mnras, 513, 6158,
  \dodoi{10.1093/mnras/stac935}

\bibitem[{{D'Orazio} {et~al.}(2013){D'Orazio}, {Haiman}, \&
  {MacFadyen}}]{Orazio2013MNRAS.436.2997D}
{D'Orazio}, D.~J., {Haiman}, Z., \& {MacFadyen}, A. 2013, \mnras, 436, 2997,
  \dodoi{10.1093/mnras/stt1787}

\bibitem[{{Duffell} {et~al.}(2020){Duffell}, {D'Orazio}, {Derdzinski},
  {Haiman}, {MacFadyen}, {Rosen}, \& {Zrake}}]{Duffell2020}
{Duffell}, P.~C., {D'Orazio}, D., {Derdzinski}, A., {et~al.} 2020, \apj, 901,
  25, \dodoi{10.3847/1538-4357/abab95}

\bibitem[{Ennoggi {et~al.}(2025)}]{Ennoggi:2025nht}
Ennoggi, L., {et~al.} 2025.
\newblock \doarXiv{2502.06389}

\bibitem[{{Farris} {et~al.}(2014){Farris}, {Duffell}, {MacFadyen}, \&
  {Haiman}}]{Farris2014}
{Farris}, B.~D., {Duffell}, P., {MacFadyen}, A.~I., \& {Haiman}, Z. 2014, \apj,
  783, 134, \dodoi{10.1088/0004-637X/783/2/134}

\bibitem[{Gold(2019)}]{gold_relativistic_2019}
Gold, R. 2019, Galaxies, 7, 63, \dodoi{10.3390/galaxies7020063}

\bibitem[{{Goldreich} \& {Tremaine}(1980)}]{Peter+80}
{Goldreich}, P., \& {Tremaine}, S. 1980, \apj, 241, 425, \dodoi{10.1086/158356}

\bibitem[{{Guti{\'e}rrez} {et~al.}(2024){Guti{\'e}rrez}, {Combi}, \&
  {Ryan}}]{Gutierrez2024}
{Guti{\'e}rrez}, E.~M., {Combi}, L., \& {Ryan}, G. 2024, arXiv e-prints,
  arXiv:2405.14843, \dodoi{10.48550/arXiv.2405.14843}

\bibitem[{{Hawley} \& {Balbus}(1995)}]{Hawley1995}
{Hawley}, J.~F., \& {Balbus}, S.~A. 1995, \pasa, 12, 159,
  \dodoi{10.1017/S1323358000020208}

\bibitem[{{Hayasaki} {et~al.}(2008){Hayasaki}, {Mineshige}, \&
  {Ho}}]{Hayasaki2008}
{Hayasaki}, K., {Mineshige}, S., \& {Ho}, L.~C. 2008, \apj, 682, 1134,
  \dodoi{10.1086/588837}

\bibitem[{{Hirose} {et~al.}(2009){Hirose}, {Krolik}, \& {Blaes}}]{Hirose2009}
{Hirose}, S., {Krolik}, J.~H., \& {Blaes}, O. 2009, \apj, 691, 16,
  \dodoi{10.1088/0004-637X/691/1/16}

\bibitem[{{Hirose} {et~al.}(2006){Hirose}, {Krolik}, \& {Stone}}]{Hirose2006}
{Hirose}, S., {Krolik}, J.~H., \& {Stone}, J.~M. 2006, \apj, 640, 901,
  \dodoi{10.1086/499153}

\bibitem[{{Ivanov} {et~al.}(1999){Ivanov}, {Papaloizou}, \&
  {Polnarev}}]{Ivanov1999}
{Ivanov}, P.~B., {Papaloizou}, J.~C.~B., \& {Polnarev}, A.~G. 1999, \mnras,
  307, 79, \dodoi{10.1046/j.1365-8711.1999.02623.x}

\bibitem[{{King} \& {Pringle}(2007)}]{King2007}
{King}, A.~R., \& {Pringle}, J.~E. 2007, \mnras, 377, L25,
  \dodoi{10.1111/j.1745-3933.2007.00296.x}

\bibitem[{Kocsis {et~al.}(2012)Kocsis, Haiman, \& Loeb}]{Kocsis:2012cs}
Kocsis, B., Haiman, Z., \& Loeb, A. 2012, Mon. Not. Roy. Astron. Soc., 427,
  2660, \dodoi{10.1111/j.1365-2966.2012.22129.x}

\bibitem[{{Kolykhalov} \& {Syunyaev}(1979)}]{Kolykhalov1979}
{Kolykhalov}, P.~I., \& {Syunyaev}, R.~A. 1979, Soviet Astronomy Letters, 5,
  180

\bibitem[{{Lai} \& {Mu{\~n}oz}(2023)}]{Lai2023}
{Lai}, D., \& {Mu{\~n}oz}, D.~J. 2023, \araa, 61, 517,
  \dodoi{10.1146/annurev-astro-052622-022933}

\bibitem[{{Lan{\v{c}}ov{\'a}} {et~al.}(2019){Lan{\v{c}}ov{\'a}}, {Abarca},
  {Klu{\'z}niak}, {Wielgus}, {S{\k{a}}dowski}, {Narayan}, {Schee},
  {T{\"o}r{\"o}k}, \& {Abramowicz}}]{Lan2019}
{Lan{\v{c}}ov{\'a}}, D., {Abarca}, D., {Klu{\'z}niak}, W., {et~al.} 2019,
  \apjl, 884, L37, \dodoi{10.3847/2041-8213/ab48f5}

\bibitem[{Lin \& Papaloizou(1979)}]{Lin+79}
Lin, D. N.~C., \& Papaloizou, J. 1979, Monthly Notices of the Royal
  Astronomical Society, 186, 799, \dodoi{10.1093/mnras/186.4.799}

\bibitem[{{Lin} \& {Papaloizou}(1986)}]{Lin1986}
{Lin}, D.~N.~C., \& {Papaloizou}, J. 1986, \apj, 309, 846,
  \dodoi{10.1086/164653}

\bibitem[{{LISA Consortium Waveform Working Group} {et~al.}(2023){LISA
  Consortium Waveform Working Group}, Afshordi, Akçay, Amaro~Seoane,
  Antonelli, Aurrekoetxea, Barack, Barausse, Benkel, Bernard, Bernuzzi, Berti,
  Bonetti, Bonga, Bozzola,
  {et~al.}}]{lisa_consortium_waveform_working_group_waveform_2023}
{LISA Consortium Waveform Working Group}, Afshordi, N., Akçay, S., {et~al.}
  2023, Waveform {Modelling} for the {Laser} {Interferometer} {Space}
  {Antenna}, \dodoi{10.48550/arXiv.2311.01300}

\bibitem[{Liu \& Shapiro(2010)}]{Liu+10}
Liu, Y.~T., \& Shapiro, S.~L. 2010, Phys. Rev. D, 82, 123011,
  \dodoi{10.1103/PhysRevD.82.123011}

\bibitem[{{Lodato} \& {Bertin}(2003)}]{Lodato2003}
{Lodato}, G., \& {Bertin}, G. 2003, \aap, 408, 1015,
  \dodoi{10.1051/0004-6361:20031045}

\bibitem[{Lodato {et~al.}(2009)Lodato, Nayakshin, King, \& Pringle}]{Lodato+09}
Lodato, G., Nayakshin, S., King, A.~R., \& Pringle, J.~E. 2009, Monthly Notices
  of the Royal Astronomical Society, 398, 1392,
  \dodoi{10.1111/j.1365-2966.2009.15179.x}

\bibitem[{{Lynden-Bell}(1969)}]{Lynden-Bell1969}
{Lynden-Bell}, D. 1969, \nat, 223, 690, \dodoi{10.1038/223690a0}

\bibitem[{Manikantan {et~al.}(2024)Manikantan, Paschalidis, \&
  Bozzola}]{Manikantan:2024giq}
Manikantan, V., Paschalidis, V., \& Bozzola, G. 2024.
\newblock \doarXiv{2411.11955}

\bibitem[{Manikantan {et~al.}(2025)Manikantan, Paschalidis, \&
  Bozzola}]{Manikantan:2025afy}
---. 2025.
\newblock \doarXiv{2504.12375}

\bibitem[{{Milosavljevi{\'c}} \& {Merritt}(2001)}]{Milosavljevi2001}
{Milosavljevi{\'c}}, M., \& {Merritt}, D. 2001, \apj, 563, 34,
  \dodoi{10.1086/323830}

\bibitem[{{Mu{\~n}oz} {et~al.}(2019){Mu{\~n}oz}, {Miranda}, \&
  {Lai}}]{Diego2019}
{Mu{\~n}oz}, D.~J., {Miranda}, R., \& {Lai}, D. 2019, \apj, 871, 84,
  \dodoi{10.3847/1538-4357/aaf867}

\bibitem[{{Narayan}(2000)}]{Narayan2000}
{Narayan}, R. 2000, \apj, 536, 663, \dodoi{10.1086/308956}

\bibitem[{{Narayan} \& {Yi}(1994)}]{Narayan1994}
{Narayan}, R., \& {Yi}, I. 1994, \apjl, 428, L13, \dodoi{10.1086/187381}

\bibitem[{{O'Neill} {et~al.}(2009){O'Neill}, {Miller}, {Bogdanovi{\'c}},
  {Reynolds}, \& {Schnittman}}]{Neill2009}
{O'Neill}, S.~M., {Miller}, M.~C., {Bogdanovi{\'c}}, T., {Reynolds}, C.~S., \&
  {Schnittman}, J.~D. 2009, \apj, 700, 859, \dodoi{10.1088/0004-637X/700/1/859}

\bibitem[{{Paczy{\'n}sky} \& {Wiita}(1980)}]{Paczy1980}
{Paczy{\'n}sky}, B., \& {Wiita}, P.~J. 1980, \aap, 88, 23

\bibitem[{Paschalidis {et~al.}(2021)Paschalidis, Bright, Ruiz, \&
  Gold}]{Paschalidis:2021ntt}
Paschalidis, V., Bright, J., Ruiz, M., \& Gold, R. 2021, Astrophys. J. Lett.,
  910, L26, \dodoi{10.3847/2041-8213/abee21}

\bibitem[{{Peters}(1964)}]{Peters1964}
{Peters}, P.~C. 1964, Physical Review, 136, 1224,
  \dodoi{10.1103/PhysRev.136.B1224}

\bibitem[{{Pringle}(1981)}]{Pringle1981}
{Pringle}, J.~E. 1981, \araa, 19, 137,
  \dodoi{10.1146/annurev.aa.19.090181.001033}

\bibitem[{Ressler {et~al.}(2025)Ressler, Combi, Ripperda, \&
  Most}]{Ressler:2024tan}
Ressler, S.~M., Combi, L., Ripperda, B., \& Most, E.~R. 2025, Astrophys. J.
  Lett., 979, L24, \dodoi{10.3847/2041-8213/ad9eb5}

\bibitem[{{Robert} {et~al.}(2018){Robert}, {Crida}, {Lega}, {M{\'e}heut}, \&
  {Morbidelli}}]{Robert2018}
{Robert}, C.~M.~T., {Crida}, A., {Lega}, E., {M{\'e}heut}, H., \& {Morbidelli},
  A. 2018, \aap, 617, A98, \dodoi{10.1051/0004-6361/201833539}

\bibitem[{{Rossi} {et~al.}(2010){Rossi}, {Lodato}, {Armitage}, {Pringle}, \&
  {King}}]{Rossi2010}
{Rossi}, E.~M., {Lodato}, G., {Armitage}, P.~J., {Pringle}, J.~E., \& {King},
  A.~R. 2010, \mnras, 401, 2021, \dodoi{10.1111/j.1365-2966.2009.15802.x}

\bibitem[{Ruiz {et~al.}(2023)Ruiz, Tsokaros, \& Shapiro}]{Ruiz:2023hit}
Ruiz, M., Tsokaros, A., \& Shapiro, S.~L. 2023, Phys. Rev. D, 108, 124043,
  \dodoi{10.1103/PhysRevD.108.124043}

\bibitem[{{Shakura} \& {Sunyaev}(1973)}]{Shakura1973}
{Shakura}, N.~I., \& {Sunyaev}, R.~A. 1973, \aap, 24, 337

\bibitem[{{S{\k{a}}dowski}(2009)}]{Sadowski2009}
{S{\k{a}}dowski}, A. 2009, \apjs, 183, 171, \dodoi{10.1088/0067-0049/183/2/171}

\bibitem[{{S{\k{a}}dowski} {et~al.}(2011){S{\k{a}}dowski}, {Abramowicz},
  {Bursa}, {Klu{\'z}niak}, {Lasota}, \& {R{\'o}{\.z}a{\'n}ska}}]{sadowski2011}
{S{\k{a}}dowski}, A., {Abramowicz}, M., {Bursa}, M., {et~al.} 2011, \aap, 527,
  A17, \dodoi{10.1051/0004-6361/201015256}

\bibitem[{{Starling} {et~al.}(2004){Starling}, {Siemiginowska}, {Uttley}, \&
  {Soria}}]{Starling2004}
{Starling}, R. L.~C., {Siemiginowska}, A., {Uttley}, P., \& {Soria}, R. 2004,
  \mnras, 347, 67, \dodoi{10.1111/j.1365-2966.2004.07167.x}

\bibitem[{{Stone} {et~al.}(1996){Stone}, {Hawley}, {Gammie}, \&
  {Balbus}}]{Stone1996}
{Stone}, J.~M., {Hawley}, J.~F., {Gammie}, C.~F., \& {Balbus}, S.~A. 1996,
  \apj, 463, 656, \dodoi{10.1086/177280}

\bibitem[{{Tanaka} \& {Menou}(2010)}]{Tanaka2010}
{Tanaka}, T., \& {Menou}, K. 2010, \apj, 714, 404,
  \dodoi{10.1088/0004-637X/714/1/404}

\bibitem[{{Thorpe} {et~al.}(2019){Thorpe}, {Ziemer}, {Thorpe}, {Livas},
  {Conklin}, {Caldwell}, {Berti}, {McWilliams}, {Stebbins}, {Shoemaker},
  {Ferrara}, {Larson}, {Shoemaker}, {Key}, {Vallisneri}, {Eracleous},
  {Schnittman}, {Kamai}, {Camp}, {Mueller}, {Bellovary}, {Rioux}, {Baker},
  {Bender}, {Cutler}, {Cornish}, {Hogan}, {Manthripragada}, {Ware},
  {Natarajan}, {Numata}, {Sankar}, {Kelly}, {McKenzie}, {Slutsky}, {Spero},
  {Hewitson}, {Francis}, {DeRosa}, {Yu}, {Hornschemeier}, \&
  {Wass}}]{LISA_white_paper}
{Thorpe}, J.~I., {Ziemer}, J., {Thorpe}, I., {et~al.} 2019, in Bulletin of the
  American Astronomical Society, Vol.~51, 77, \dodoi{10.48550/arXiv.1907.06482}

\bibitem[{Tiwari {et~al.}(2025)Tiwari, Chan, Bogdanovi\'c, Jiang, Davis, \&
  Ferrel}]{Tiwari:2025imm}
Tiwari, V., Chan, C.-H., Bogdanovi\'c, T., {et~al.} 2025.
\newblock \doarXiv{2502.18584}

\bibitem[{{Turpin} \& {Nelson}(2024)}]{Turpin2024}
{Turpin}, G.~A., \& {Nelson}, R.~P. 2024, \mnras, 528, 7256,
  \dodoi{10.1093/mnras/stae109}

\bibitem[{{Valli} {et~al.}(2024){Valli}, {Tiede}, {Vigna-G{\'o}mez}, {Cuadra},
  {Siwek}, {Ma}, {D'Orazio}, {Zrake}, \& {de Mink}}]{Valli2024}
{Valli}, R., {Tiede}, C., {Vigna-G{\'o}mez}, A., {et~al.} 2024, arXiv e-prints,
  arXiv:2401.17355, \dodoi{10.48550/arXiv.2401.17355}

\bibitem[{{Wang} {et~al.}(2023){Wang}, {Bai}, \& {Lai}}]{Wang2023}
{Wang}, H.-Y., {Bai}, X.-N., \& {Lai}, D. 2023, \apj, 943, 175,
  \dodoi{10.3847/1538-4357/acac77}

\bibitem[{Westernacher-Schneider {et~al.}(2022)Westernacher-Schneider, Zrake,
  MacFadyen, \& Haiman}]{westernacher-schneider_multi-band_2022}
Westernacher-Schneider, J.~R., Zrake, J., MacFadyen, A., \& Haiman, Z. 2022,
  Physical Review D, 106, 103010, \dodoi{10.1103/PhysRevD.106.103010}

\end{thebibliography}
\bibliographystyle{aasjournal}

\end{document}